\documentclass[a4paper,11pt]{article}
\usepackage{pos}
\usepackage{wrapfig}
\usepackage{enumitem}
\usepackage{hyperref}
\usepackage{lineno}

\title{Highlights from the Pierre Auger Observatory}
\ShortTitle{Highlights from the Pierre Auger Observatory}

\author*[a,b]{Francesco Salamida}
\onbehalf{for the Pierre Auger Collaboration$^c$}

\affiliation[a]{ Università dell’Aquila, Dipartimento di Scienze Fisiche e Chimiche, L’Aquila, Italy}
\affiliation[b]{ INFN Laboratori Nazionali del Gran Sasso, Assergi (L’Aquila), Italy}

\affiliation[c]{Observatorio Pierre Auger, Av.\ San Mart{\'\i}n Norte 304, 5613 Malarg\"ue, Argentina\\
Full author list: \normalfont{\url{https://www.auger.org/archive/authors\_icrc\_2023.html}}}

\emailAdd{spokespersons@auger.org}

\abstract{The Pierre Auger Observatory is a unique facility designed to study ultra-high energy cosmic rays, with energies up to 10$^{20}$~eV and beyond. The Observatory is located in Argentina and comprises more than 1600 water Cherenkov detectors spread over an area of 3000 square kilometers overlooked by Fluorescence detectors. The first phase of the Observatory's data-taking began in 2004 and continued until the end of 2021. In this contribution, the results from the Phase~I data analysis of the Pierre Auger Observatory are presented. They include, among others, measurements of the cosmic-ray energy spectrum, composition, and arrival direction anisotropy. The Phase~I results from the Pierre Auger Observatory provide major advances in the understanding of the ultra-high energy cosmic ray phenomena and lay the foundation for second-phase studies with the upgraded AugerPrime detector. The status of the AugerPrime upgrade and its performance will be also discussed.}

\ConferenceLogo{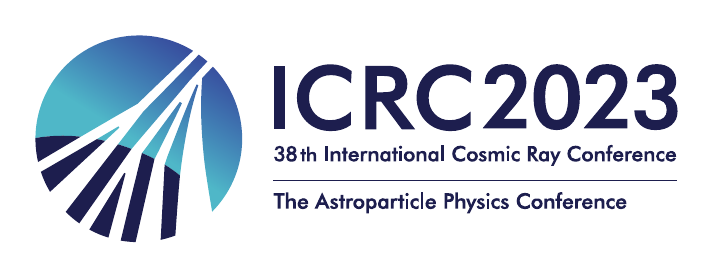}

\FullConference{%
38th International Cosmic Ray Conference (ICRC2023)\\
26 July -- 3 August, 2023\\
Nagoya, Japan}

\begin{document}
\maketitle

\section{Introduction}

The study of ultra-high energy cosmic rays (UHECRs) above $10^{17}$~eV opens a window of opportunity in the understanding of their origin. More than 100 years after the discovery of cosmic rays by Victor Hess, their sources are largely unknown. The main reason is that, regardless of the angular resolution with which they are revealed, their direction of arrival does not appear to be largely correlated with the positions of known sources. The application of the most intuitive method of association to the source by direction, analogous to photons at different frequencies of observation is in fact made very difficult by the presence of magnetic fields in our Galaxy (of the order of
$10^{-6}$~G) and in extragalactic space (of the order of $10^{-9}$~G). Although we are yet to develop a comprehensive theory that can elucidate the production sites of UHECRs, the mechanisms accelerating atomic nuclei to such extraordinary energies, and the precise composition of these cosmic rays, significant progress has been made in recent years by exploiting data obtained from the Pierre Auger Observatory. The most extensive dataset of UHECR events ever collected has significantly advanced our comprehension of their characteristics, revealing a global perspective that appears considerably more intricate than what was previously expected. A striking example is the results on the primary composition of cosmic rays, which show that masses become heavier above energy almost coincident with the ankle position, with mass groups replacing each other so that the flow of all particles is dominated by only one specific mass group in each energy range. This completely ruled out the paradigm according to which a proton-only composition at the highest energies provided a unique explanation for the presence of both the ankle~\cite{dipmodel} and the suppression~\cite{GZK1,GZK2} in the spectrum and was based on the assumption that the mechanisms of extreme acceleration operated on hydrogen nuclei being the most abundant. A key role in understanding these results is attributed to models for hadronic interactions, which, based on the latest findings from particle accelerators, must be extrapolated to extreme energies. In this region, UHECRs are the only instruments available for physics, and measurements at the Pierre Auger Observatory show an excess in the number of muons of between 20\% and 30\% compared to current interaction models. To improve composition information, the Pierre Auger Observatory is currently undergoing an upgrade, namely AugerPrime~\cite{ICRC_convenga, ICRC_anastasi}, which will transition it from Phase~I to Phase~II of its life. This contribution will highlight the status of the upgrade and the most recent results achieved using the Phase~I data of the Observatory and presented at this conference.

\section{The Pierre Auger Observatory and the AugerPrime upgrade}

In 1992 James Cronin and Alan Watson proposed the idea of building a surface detector of giant dimensions and thus attempting to answer the fundamental questions about the origin of UHECRs. The basic design of the Pierre Auger Observatory~\cite{augerNIM} is made of an array consisting of about 1600 surface detectors (SD) arranged on a triangular grid and spaced 1500~m apart. Additionally, two nested arrays of the same kind with 750~m and 433~m spacing were utilized to lower the energy threshold down to 63~PeV. At a later stage, a 17~km$^2$ area was instrumented with the Auger Engineering Radio Array (AERA). The fluorescence detector (FD) consists of 24 telescopes spanning 3-30 degrees elevation and grouped in 4 sites, overlooking the array to capture the fluorescence light produced by showers in the atmosphere. Three newer telescopes of the same type are installed close to the denser arrays; they can be tilted to cover elevations up to 60 degrees thus allowing us to measure showers at lower energies.

\begin{figure}
     \centering
      \begin{tabular}{cc}
        \includegraphics[height=0.35\textheight]{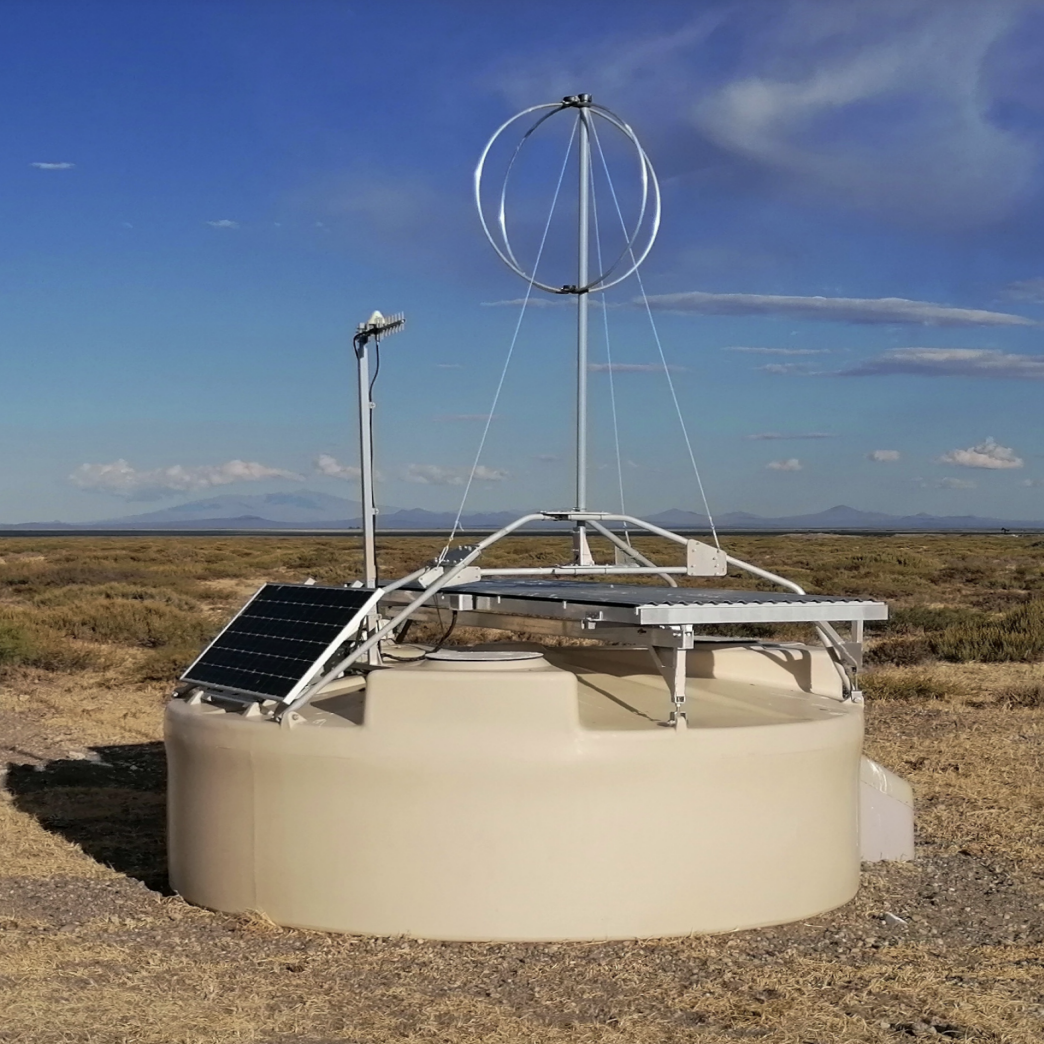} &
        \includegraphics[height=0.35\textheight]{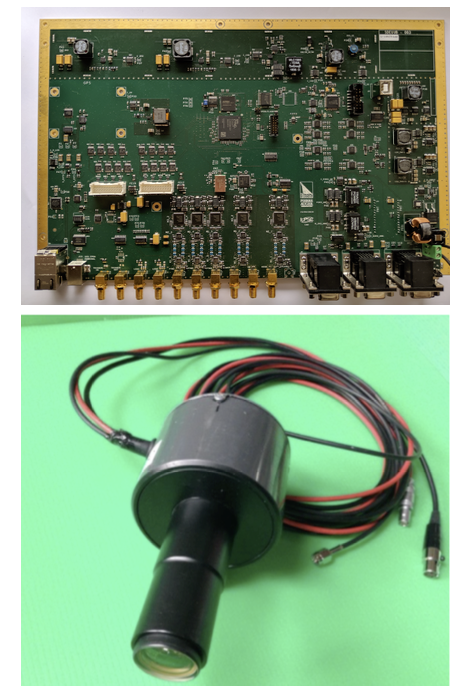}
     \end{tabular}
     \caption{Left: a surface station equipped with all the AugerPrime
       detectors. The image shows the SSD positioned on top of the
       WCD, accompanied by the radio detector on top of the
       SSD. Right: in the upper panel there is a picture of the UUB
       and in the lower one a picture of the sPMT. }
     \label{fig:augerprime}
 \end{figure}

\subsection{AugerPrime}

 water Cherenkov detector (WCD) of the Observatory, of a scintillator-based surface detector~(SSD)~\cite{ICRC_cataldi} each covering an area of about 2~m$^2$ and enclosed in an aluminum box. The light, produced in the scintillator bars, is collected and propagated along the WLS fibers and read out by the same PMT in the central part of the module. Furthermore, a radio detector~(RD)~\cite{ICRC_pawlosky} will be added. It consists of a dual-polarized antenna, one of the antenna rings is aligned parallel to the orientation of the Earth’s magnetic field, while the second is perpendicular to it. The antenna is a Short Aperiodic Loaded Loop Antenna with a diameter of 122~cm to be tailored at the frequency range of 30-80~MHz. This modified configuration aims to enhance the capability to distinguish between the muonic and electromagnetic components of the shower on an event-by-event basis leveraging the distinct responses of these detectors. Additionally, the SSD and RD will provide complementary information too: the SSD is optimized for detecting more vertical showers, while the RD is fine-tuned for showers at higher zenith angles. As part of the AugerPrime project, a small 1-inch Hamamatsu R8619 photomultiplier (sPMT)~\cite{ICRC_anastasi} has been installed alongside the existing three large PMTs (LPMTs) at each surface station, to largely reduce the occurrence of saturated signals in the stations closest to the shower axis. A photo of the SPMT is shown in the lower right panel of figure~\ref{fig:augerprime}. Furthermore, underground muon detectors (UMDs)~\cite{ICRC_dejesus} are being strategically placed in the denser areas, where the 433~m and 750~m arrays are located, to directly measure the muon component of the cosmic ray showers to understand the significant discrepancy between models and data appearing in the average muon scale~\cite{ICRC_whisp}. UMDs are plastic scintillator strips buried 2.3~m deep to ensure that the electromagnetic component of showers is largely absorbed while vertical muons with energy above 1 GeV can reach the buried detectors. The upgraded station configuration is depicted on the left panel of figure~\ref{fig:augerprime}, showcasing the installation of SSD and RD on top of the station. Furthermore, there has been an (instead of the UB's 40~MHz) and 12~bit accuracy which can receive and manage the signals from the various components of the upgraded detector. The backward compatibility with the data recorded in Phase~I is obtained by filtering and downsampling the traces to emulate the current triggers in addition to any new ones.
Furthermore, the data acquisition software, both at the level of the individual detector stations and the central data acquisition system, was also updated~\cite{ICRC_sato}. The changes were necessary to handle the new multi-hybrid data from AugerPrime and to cope with surface detection stations with different hardware configurations operating simultaneously in the array during the transition phase. In July 2023, the installation phase of SSDs, sPMTs, and UUBs on the entire array was completed. After the due commissioning period, full efficiency data taking, i.e.\ Phase~II, is expected to start in 2024 and is foreseen to add 10 more years of data. The installation of the UMD detectors and radio antennas on the array stations remains to be completed. According to our timetable, the installation activities are scheduled to be completed around mid-2024.

 \subsection{AugerPrime performances and first data} 
 
\begin{figure}
     \centering
      \begin{tabular}{cc}
        \includegraphics[width=0.57\textwidth]{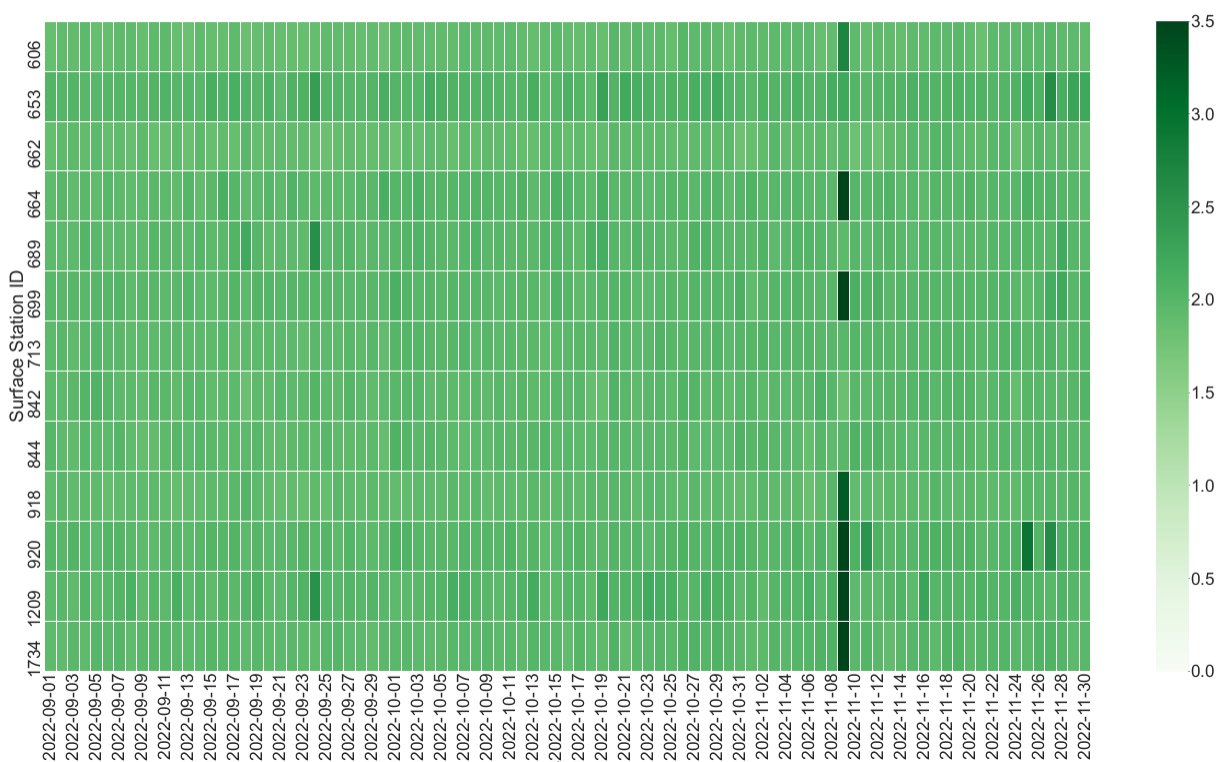} &
        \includegraphics[width=0.4\textwidth]{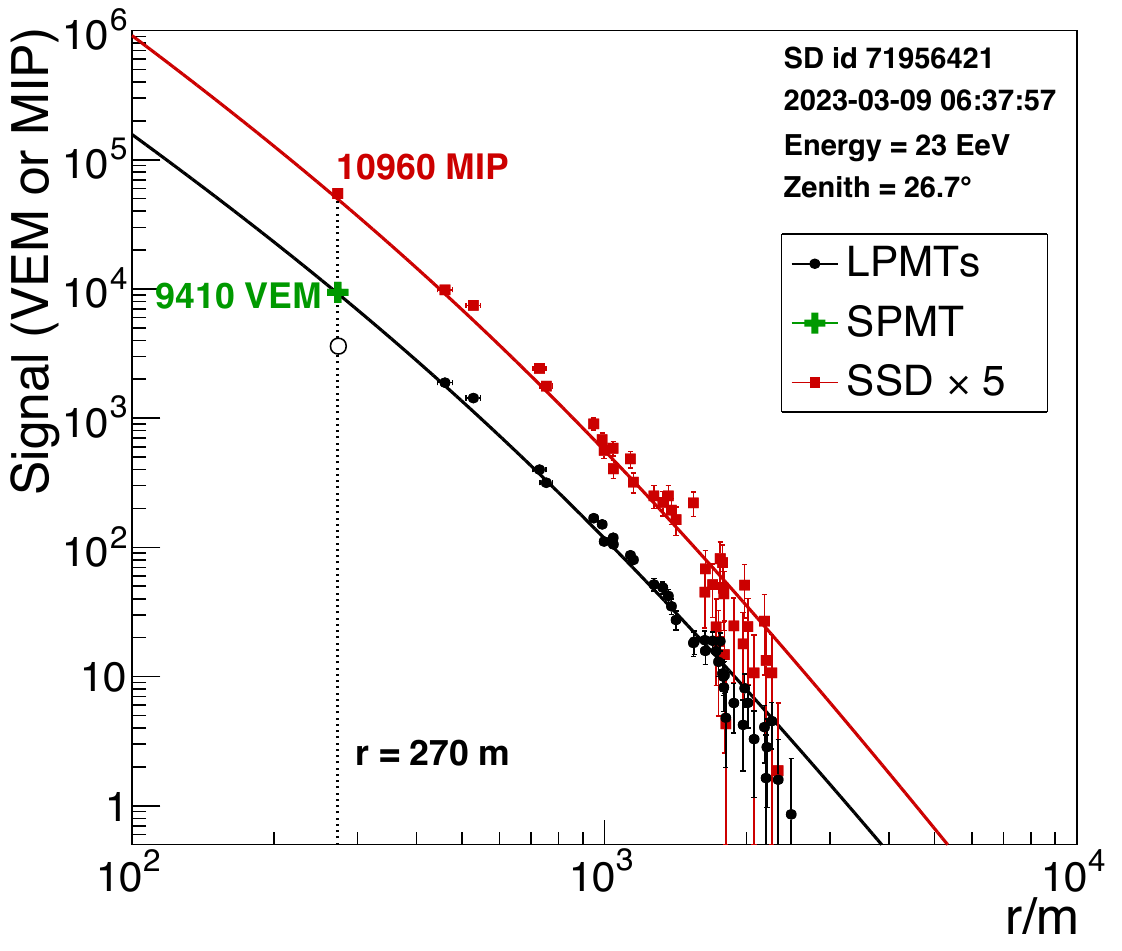}
      \end{tabular}
      \caption{Left: map of daily mean values of the baseline RMS. Inhomogeneities resulting in darker vertical lines correspond to noise due to lightning storms in the array. Right: high-energy event collected by AugerPrime detectors and detected with the WCDs (black dots) and SSDs (red squares). In the station nearest to the shower core the LPMTs are saturated, thus the SPMT signal is used (green cross). The SSD signals and SSD LDF are multiplied by a factor of 5 to be shown on the same plot. }      
     \label{fig:primeperformances}
 \end{figure}

\begin{figure}
     \centering
      \begin{tabular}{cc}
        \includegraphics[width=0.36\textwidth]{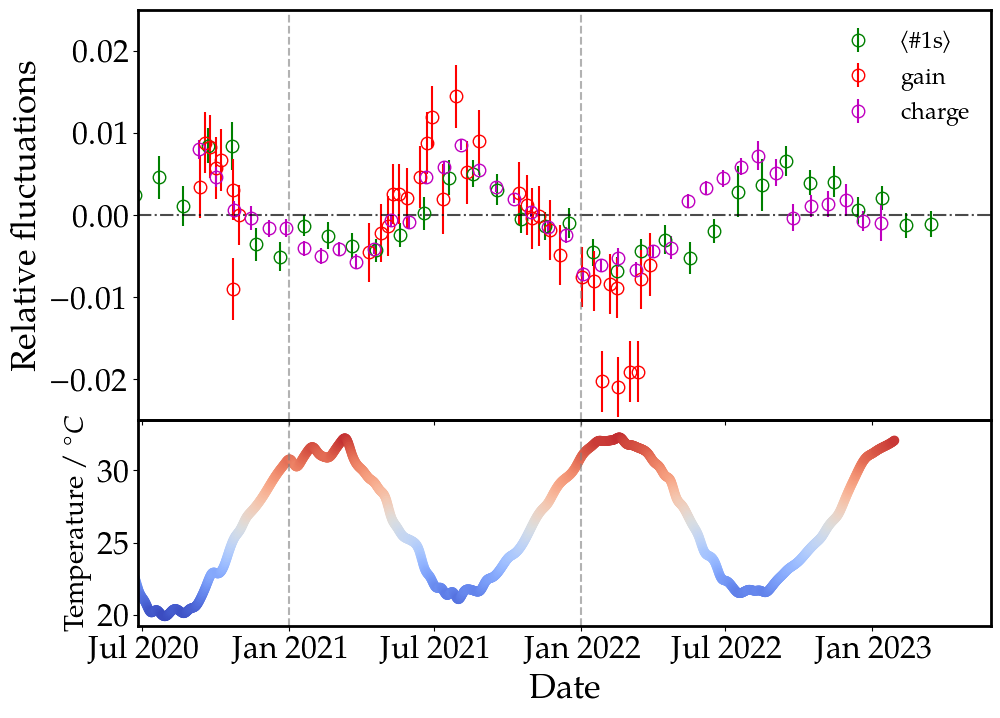} &
        \includegraphics[width=0.63\textwidth]{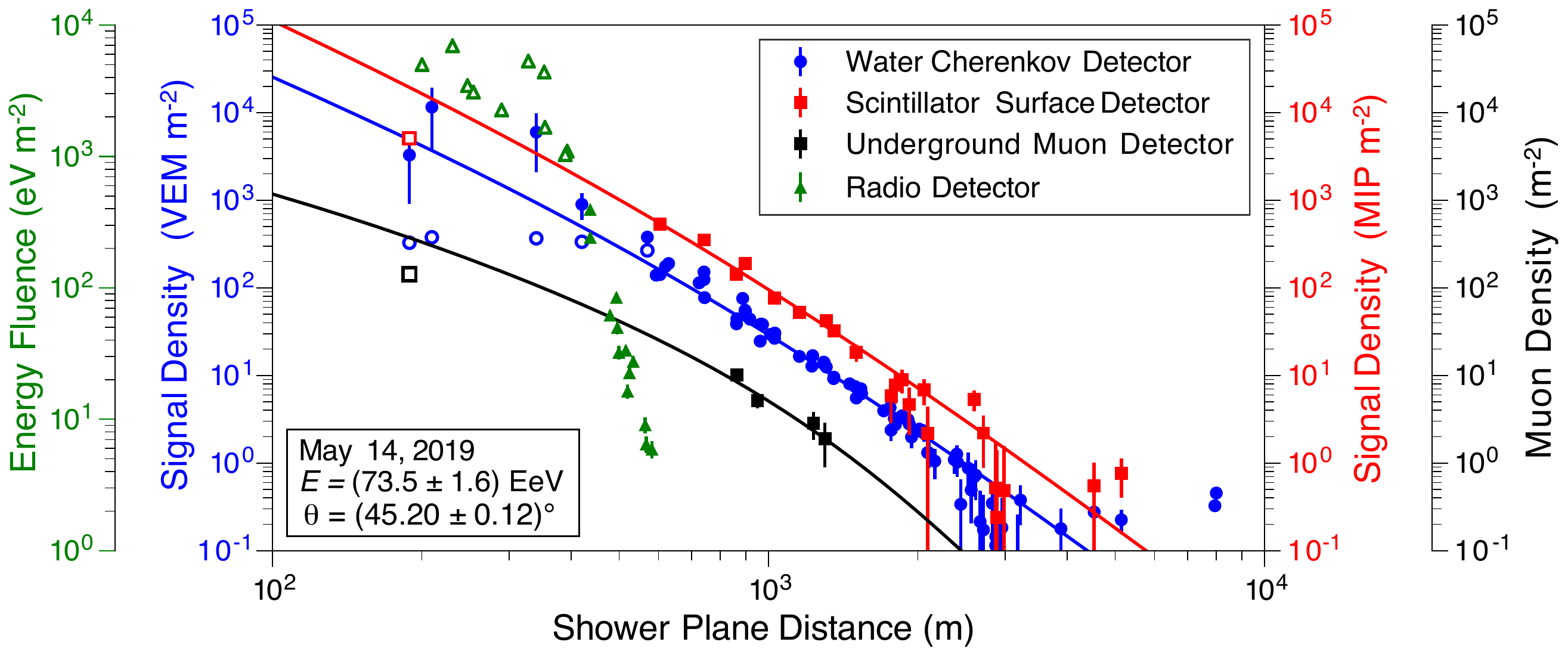} 
      \end{tabular}
     \caption{Left: seasonal fluctuations of the UMD gain and its impact on acquisition parameters. The stability of the acquisition is within $\sim 1$~\%. Right: one of the earlier multi-hybrid events collected by AugerPrime detectors.}
     \label{fig:primeperformances2}
   \end{figure}

 While implementing the new electronic systems, there was no interruption of the data acquisition. We also performed updates to the central data acquisition software (CDAS) and data analysis pipelines for AugerPrime. We continuously monitored and analyzed the collected data to evaluate the ongoing performance of the detection stations. The parameters tracked during the deployment include the RMS of the ADC signal baseline as an indicator of electronic noise, the VEM charge, the MIP charge~\cite{ICRC_convenga}, and the $\beta$ factor of each PMT which quantifies the stability of this gain factor over time. In particular, figure~\ref{fig:primeperformances}~(left panel) shows the evolution of the daily mean baseline RMS expressed in ADC counts for a subsample of detectors~(on the vertical axis) in the array during three months of data taking. The uniformity in time indicates the stability of the detector's operations within the expected seasonal fluctuations. Darker vertical lines point out periods of bad weather conditions (e.g.\ lightning) and are currently the object of ongoing studies to rule out time intervals unsuitable for physics analysis. The potential of the enhanced detector was already evident as soon as the first events were acquired. As already described, the small PMT extends the dynamic range of the WCD signals, allowing the recovery of any saturated stations as shown in figure~\ref{fig:primeperformances}~(right panel). In addition, the event in the figure shows how the SSD design allows for a complementary measurement to the upgraded WCD for signals close to the shower core. Although the installation phase of the UMD and RD~\cite{ICRC_pawlosky} is still in progress, we carry out a monitoring of the performance of the detectors already deployed. In particular, we can see in figure~\ref{fig:primeperformances2} how the relative fluctuations for the signals acquired by the UMD SiPMs are at the $\pm 1$~\% level and, as expected, correlated with temperature variations. We can even better understand the high potential of the new data from the enhanced observatory by looking at figure~\ref{fig:primeperformances2}~(right panel), where a multi-hybrid event with simultaneous signal in the WCD, SSD, UMD, and RD is shown. The simultaneous analysis of the signals from the different detectors will make it possible in the near future to obtain an estimate of the primary mass on an event-by-event basis by separating the muon and electromagnetic components.
 
\subsection{Novel detectors and analysis methods}

The Pierre Auger Observatory continually demonstrates its vitality through enhancements in both data analysis methods and instrument upgrades. This commitment ensures that the Observatory remains at the forefront of UHECR physics enhancing its precision and promising the potential for groundbreaking discoveries in the field. In particular, a novel technique for the absolute calibration of the fluorescence telescopes has been developed~\cite{ICRC_schafer} (i.e.\ XY scanner). A scan (with ${\sim}1700$ positions) of the complete aperture of a telescope is performed by a uniformly emitting, absolutely calibrated light source mounted on a rail system. A dedicated setup with a NIST-calibrated photodiode is used to calibrate the light source. This new method currently has a 6\% uncertainty, which is expected to be reduced to 4\% by improving the configuration for calibrating the light source.
A preliminary comparison between the results of the XY scanner and those obtained by another independent method based on night sky background (NSB) measurements shows good agreement~\cite{ICRC_segreto}. 
Another striking example is the use of the stereo energy balance technique based on air shower events observed simultaneously from at least two FD sites, where it is expected that along the segment of the shower track common to the views from both sites, the reconstructed energy deposit is consistent. The method allowed us to improve the systematic uncertainty on the estimation of the vertical aerosol optical depth~(VAOD), which results in an updated systematic uncertainty, in the energy range $3 \times 10^{18} - 10^{20}$~eV, equal to 2\%~-~4\% on energy and 2~g~cm$^{-2}$~-~4g~cm$^{-2}$ on X$_{\textrm{max}}$.
Other developments include i) new analysis techniques based on deep neural networks (DNN) for the primary mass identification with the SD and the SSD~\cite{ICRC_langner}, SD energy estimation~\cite{ICRC_ellwanger}, and muon number reconstruction with the SD~\cite{ICRC_hahn}; ii) the development of a new method that, by requiring a coincidence between the WCD and the SSD, will ensure a high-precision calibration of the water-based Cherenkov detectors for at least another 15 years of operation, irrespective of the further ageing of its components; iii) improvement of the fit of the longitudinal profile of FD events~\cite{ICRC_bellido}, to cope with deeper X$_\textrm{max}$ events; iv) cloud height reconstruction and study of the horizontal homogeneity of the atmosphere using elastic lidars~\cite{ICRC_pallotta}; 

 \section{Physics results}
 The Phase~I data of the Pierre Auger Observatory is exceptionally compelling and constitutes the world's largest database of events at extreme energies. In the following, we will provide an update on the most significant results presented at this conference by the Pierre Auger Collaboration.

\subsection{Extension of the flux measurements down to $6 \times 10^{16}$~eV}

 \begin{figure}
     \centering
      \begin{tabular}{cc}
        \includegraphics[width=0.48\textwidth]{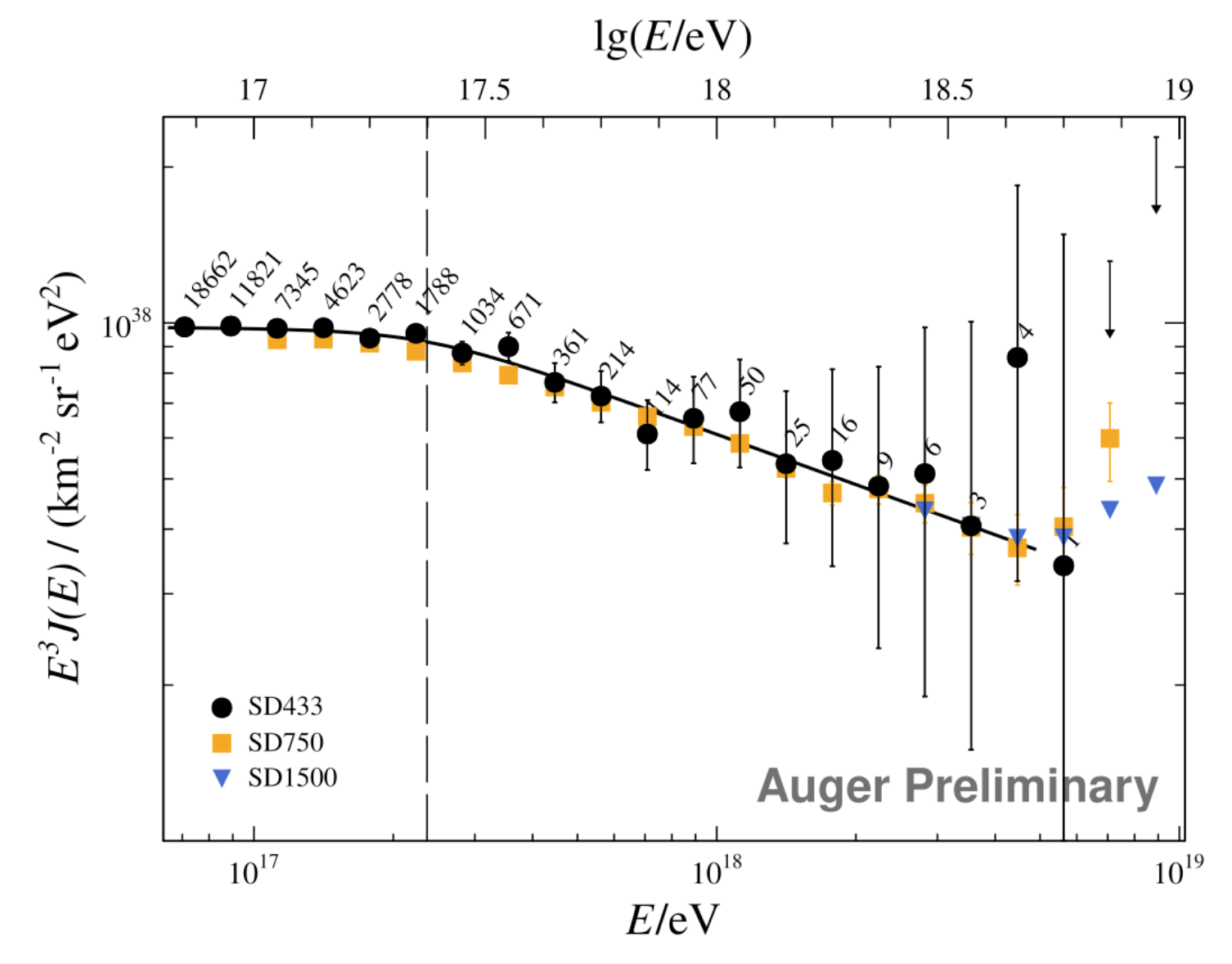}& 
        \includegraphics[width=0.48\textwidth]{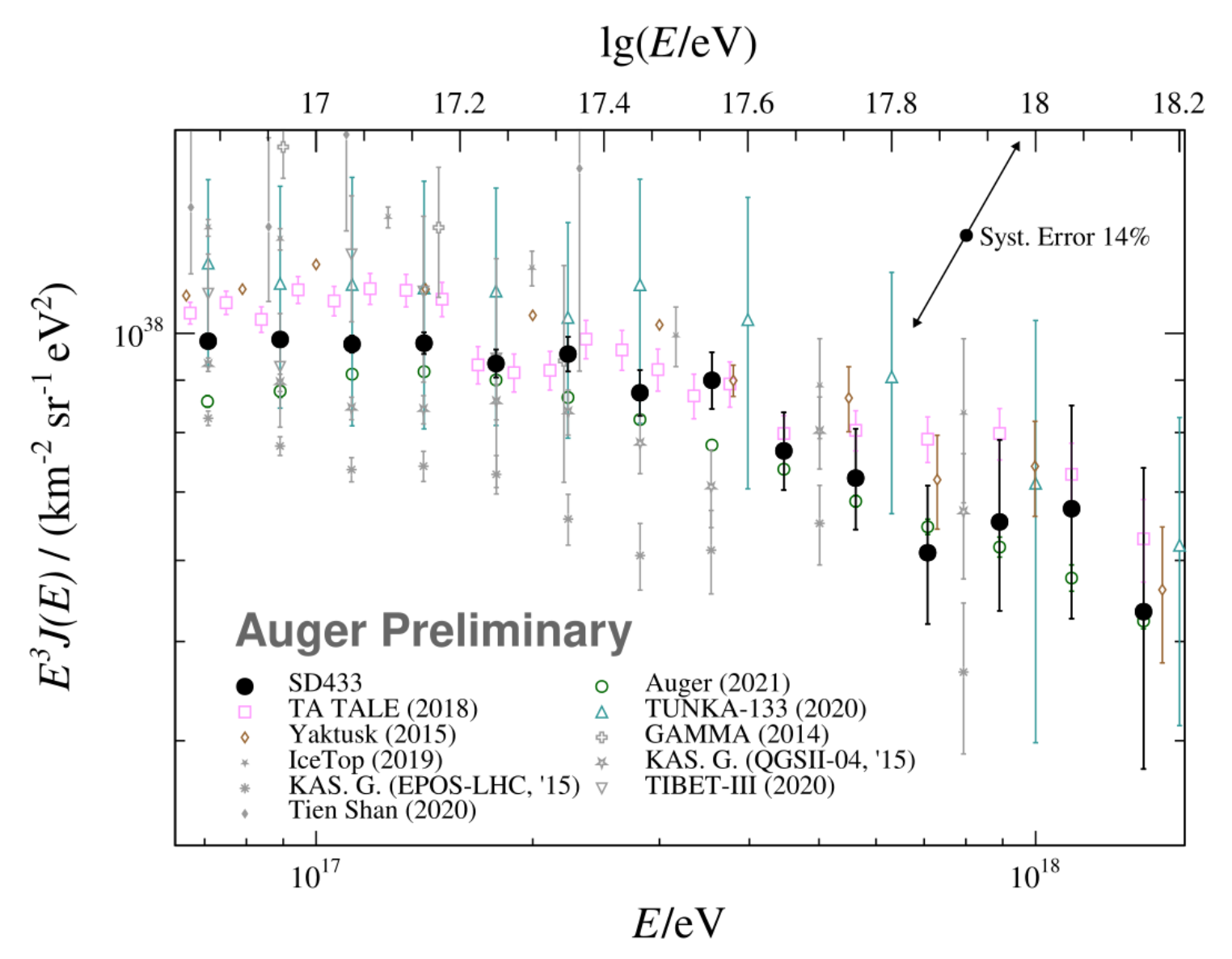} \end{tabular}
     \caption{ 
     Left: unfolded cosmic ray spectrum measured with the SD433 together with those from the 750~m array~\cite{auger_spec_eurph} and the 1500~m~\cite{auger_spec_prd} measurements. The data are fitted with a broken power law model with soft transitions and the position of the second knee is indicated with a dashed vertical line. The error bars and upper limits are also shown. Right: SD433 spectrum scaled by E~$^3$. Auger combined spectrum~\cite{ICRC_novotny} and the measurements of other experiments around the second knee~(references in \cite{auger_spec_eurph}). Colors represent experiments that have established their energy scale through calorimetric measurements.}
     \label{fig:auger_spec}
 \end{figure}

 In more than 20 years of operation, the Pierre Auger Observatory has produced results of considerable scientific relevance, including, among others, the measurement of the UHE cosmic ray flux in unprecedented detail~\cite{auger_spec_prl,auger_spec_prd,auger_spec_eurph}. The enormous exposure accumulated by the Pierre Auger Observatory (e.g.\ about 80000 km with the SD only) allows the measurement of the energy spectrum of cosmic rays with the 1500~m and the 750~m spacing arrays at energies above 10$^{17}$~eV. In addition, the measurements made with the Fluorescence detector in the hybrid mode allow an almost independent measurement in the region above 10$^{18}$~eV and with the Cherenkov mode low to about 10$^{16}$~eV.   
 We have verified with extraordinary accuracy the presence of the cut-off at $(4.7 \pm 0.3 \pm 0.6 ) \times 10^{19}$~eV, the ankle feature at  $(4.9 \pm 0.1 \pm 0.8) \times 10^{18}$eV and the presence of the second knee and, for the first time, a new structure called the \emph{instep} has been identified at $1.4 \times 10^{19}$~eV. In addition, it has been possible to independently measure the position of the second knee at  $(2.3 \pm 0.5 \pm 0.35) \times 10^{17}$~eV in agreement within 5\% with the previous measurements of the 750~m array using the 433~m  spacing array~\cite{ICRC_brichetto}. The cosmic ray spectrum reconstructed with the 433~m array is shown in figure~\ref{fig:auger_spec}~(right panel). The data were fitted with a  broken power law model with smooth transitions. The position of the second knee is indicated by the dashed line and the error bars and upper limits correspond to the statistical uncertainty given by the Feldman-Cousins intervals with a 90\% confidence level. The other SD-based measures are also shown for comparison. Our measurement is also consistent within the uncertainties with the results previously reported by Auger where the spectrum around the second knee was obtained by extrapolating the 750~m spectrum below $10^{17}$~eV using the Cherenkov dominated FD events~\cite{ICRC_novotny}. Figure~\ref{fig:auger_spec}~(right panel) shows the SD433 spectrum along with the measurements of other experiments and the previous Auger report and also demonstrates the primary role of the Pierre Auger Observatory in the energy region where the transition between Galactic and Extragalactic cosmic rays is expected to occur. 

 \subsection{Mass composition}
  \begin{figure}
     \centering
      \begin{tabular}{cc}
        \includegraphics[width=0.45\textwidth]{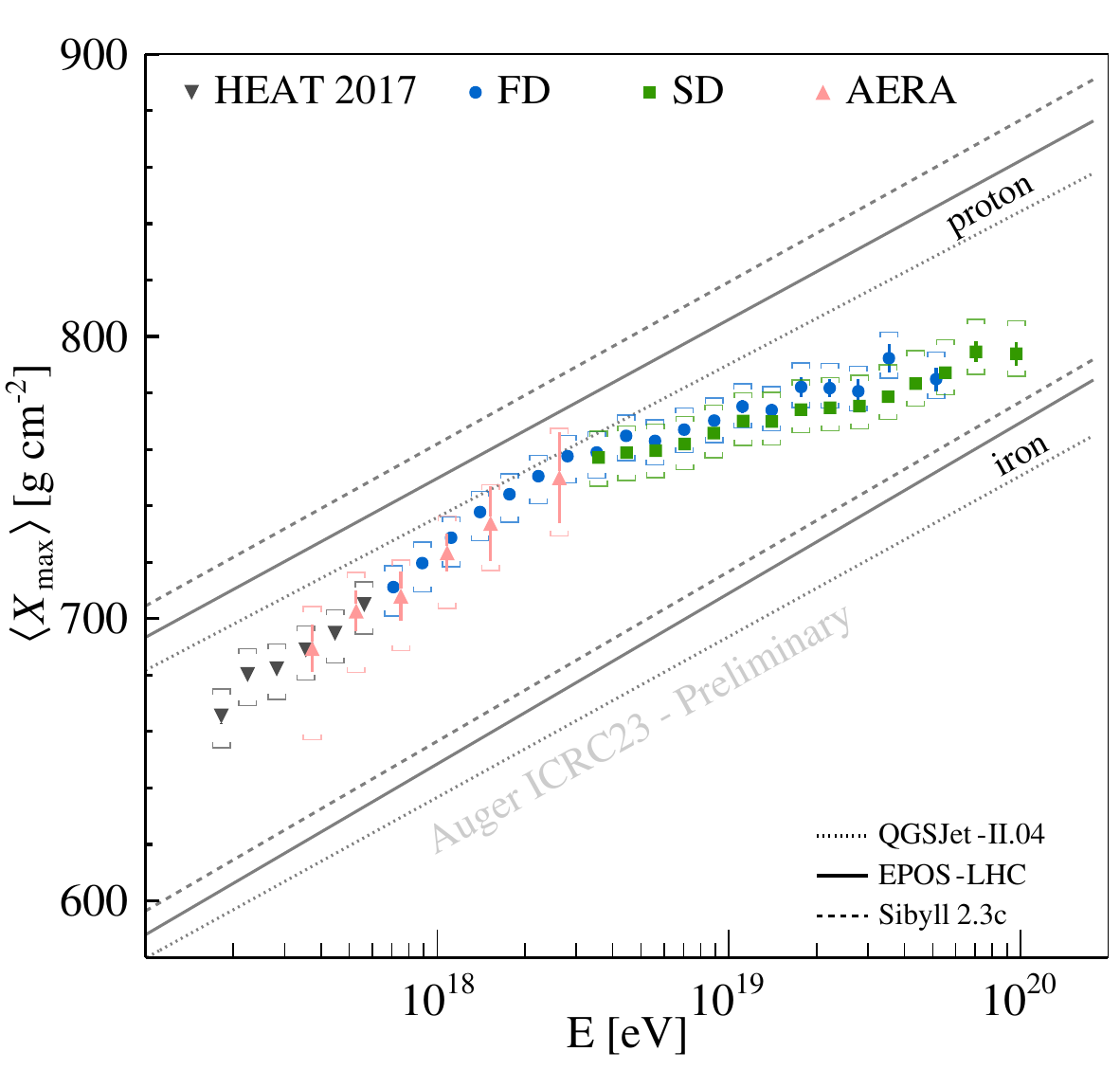} &
        \includegraphics[width=0.45\textwidth]{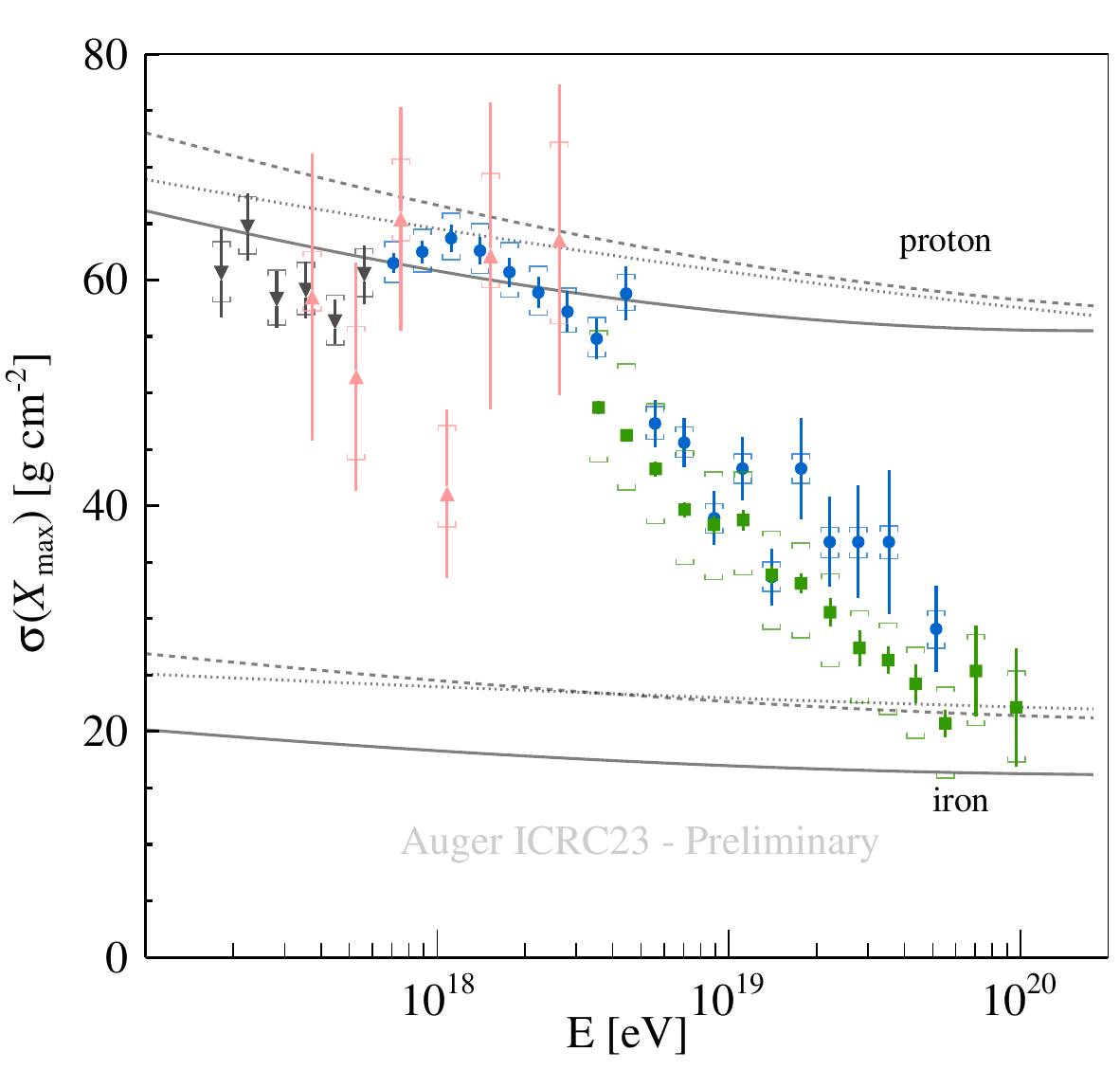}
     \end{tabular}
     \caption{The first~(left panel) and second (right panel) moments of X$_{\textrm{max}}$ distributions measured with the FD, the SD, AERA, and HEAT using Phase~I data.}
     \label{fig:auger_xmax}
   \end{figure}
   
   The depth of the cosmic ray maximum in the atmosphere provides, to the best of current knowledge, the most reliable mass-sensitive observable for the study of UHECRs' primary composition. The mean of X$_{\textrm{max}}$ is in fact directly related to the logarithmic mass of the primary particle, while the $\sigma(\textrm{X}_{\textrm{max}})$is a combination of the intrinsic shower-to-shower fluctuations and the dispersion of masses in the primary beam. The general trend of mass as a function of energy can thus be extracted by observing the evolution of the first two moments of X$_{\textrm{max}}$. Following Auger's first results~\cite{auger_mass1, auger_mass2, auger_mass3} based on the analysis of the events detected by the FD and SD, alternative analyses have been developed to estimate the depth of the shower maximum by additionally exploiting the SD detector alone, the AERA radio array, and HEAT. In particular, FD observations allow the most accurate reconstruction of X$_{\textrm{max}}$, but are only possible on dark nights and in good weather. However, the SD array has a duty cycle close to 100\%, allowing UHECRs to be measured with a higher statistic. Unfortunately, X$_{\textrm{max}}$
   cannot be observed directly with SD, but is nevertheless incorporated into the time structure of the particle footprint, which makes its reconstruction difficult. Possibilities are the use of observables, such as the signal rise time, to obtain accurate conclusions on the average variation of the composition with energy~\cite{SD_comp, ICRC_peixoto}.  On the other hand, recent advances in deep learning and associated techniques based on deep neural networks open up new possibilities for improving reconstructions in astroparticle physics~\cite{DNN_erdmann}. In particular, to reconstruct the depth of the shower maximum, the time structure of the signals measured at each SD station is exploited with deep neural networks. Both arrival times and measured signal traces are used as input for the algorithm comprising convolutional neural networks~(CNN) and long short memory networks~(LSTM)~\cite{DNN_general}. After the algorithm is cross-calibrated with the fluorescence detector observations, it is used on the SD dataset allowing for increased statistics and extended energy range compared to FD-only measurements. The most up-to-date summary~\cite{ICRC_mayotte} of the latest X$_{\textrm{max}}$ measurements performed with FD~\cite{ICRC_fitoussi}, SD~\cite{ICRC_glombitza}, AERA~\cite{AERA_mass} and HEAT~\cite{HEAT_mass} is shown in figure~\ref{fig:auger_xmax} together with the expectations for pure proton and iron primaries and for different hadronic interaction models.

   \begin{figure}
     \centering
     \begin{tabular}{cc}
       \includegraphics[width=0.46\textwidth]{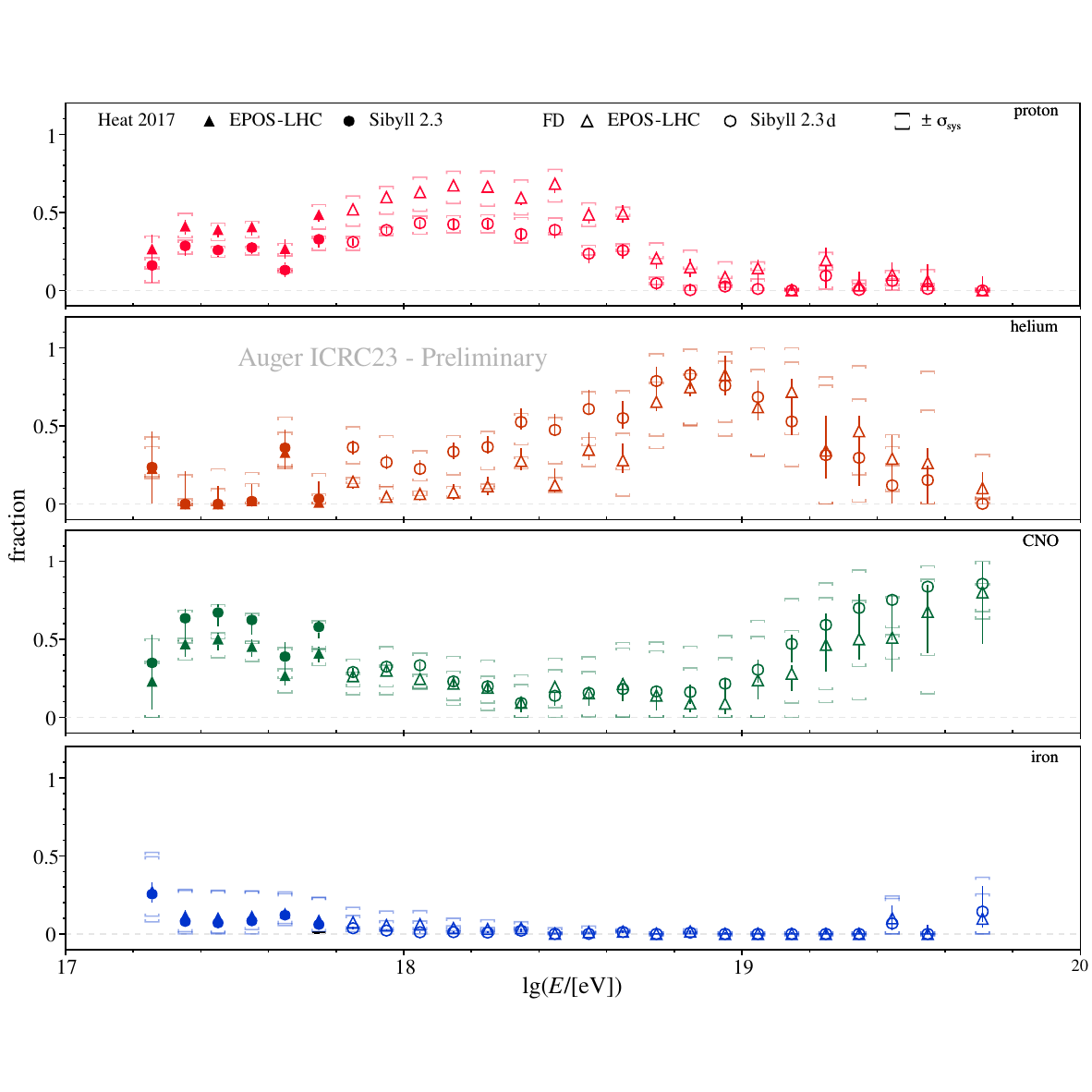} 
       &
       \includegraphics[width=0.44\textwidth]{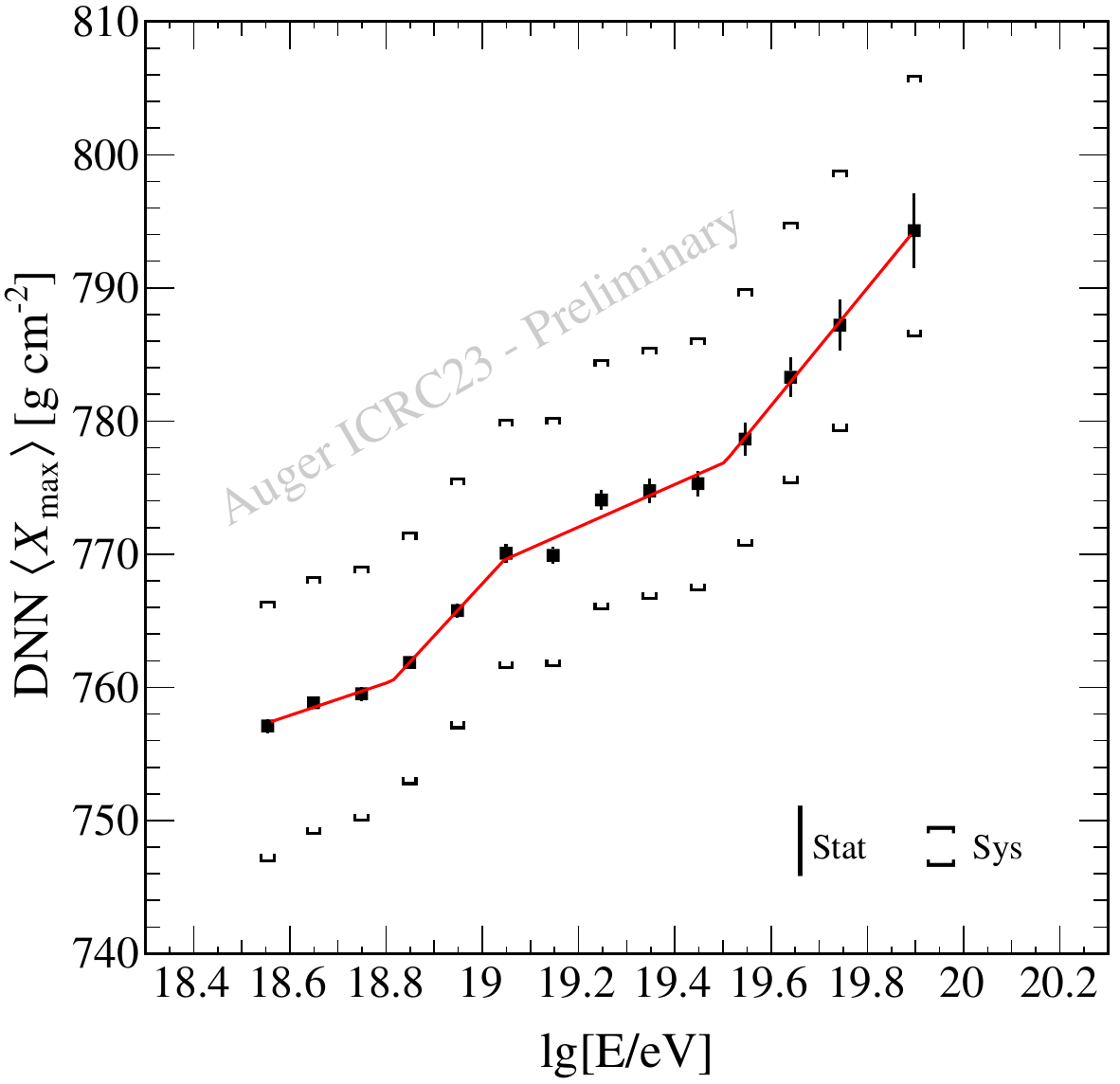}
     \end{tabular}
     \caption{Left: fits of the fractional mass composition of the UHECR flux derived from HEAT and FD X$_{\textrm{max}}$ data.  Fractions are extracted fitting parameterizations of the X$_{\textrm{max}}$  distributions for p, He, CNO, and Fe generated using different hadronic interaction models. Right: average shower maximum as determined using the DNN
       as a function of energy. Both
       statistical and systematic errors are indicated.}
     \label{fig:DNN_xmax}
 \end{figure}
 
 The agreement between different methods for the $\left<\textrm{X}_{\textrm{max}}\right>$ is within the systematic uncertainties despite the different X$_{\textrm{max}}$ scale due to the non-synchronous update of reconstruction methods. The discrepancy between the FD and SD results, visible in $\sigma(\textrm{X}_{\textrm{max}})$ instead, could arise from either remaining model-related dependencies in the SD result or potential statistical effects in the FD one. Nevertheless, it's crucial to highlight how employing SD enables us to expand X$_{\textrm{max}}$ measurements over an additional half a decade in energy. In general, Auger results indicate a UHECRs flux above $10^{17}$~eV composed of ionized atomic nuclei, with the average mass decreasing to the lightest point around $3 \times 10^{18}$~eV, and then increasing significantly. For much of the energy range, UHECRs appear to have a mixed composition, however, above a few EeV, they become increasingly more pure. To better understand this, the observed distributions of X$_{\textrm{max}}$ can be fit~\cite{ICRC_olena} with model templates of different primary mass groups and estimate how much each group contributes to the overall flux. 
The results of this process~(see left panel of figure ~\ref{fig:DNN_xmax}) shows that iron is almost entirely absent from the flux between $10^{18.4}$\,eV and $10^{19.4}$\,eV; protons are a minor component above the ankle and become rare at the highest energies; a substantial mixing of different mass groups is evident at all energy levels except the most elevated ones. In particular, the results obtained with SD are of special interest. Indeed, besides showing for the first time the possibility of reconstructing $\textrm{X}_{\textrm{max}}$ event-by-event with DNN methods, they confirm that the evolution of $\left<\textrm{X}_{\textrm{max}}\right>$ cannot be described by a constant elongation rate. Using a more complex model,
 a distinctive pattern emerges indicating the presence of three breakpoints in the elongation rate, as illustrated in the left panel of figure~\ref{fig:DNN_xmax}. In this model, the elongation rate  increases from 12~$\pm$~5~g~cm$^{-2}$/decade of energy to 39~$\pm$~9~g~cm$^{-2}$/decade at an energy of (6.5~$\pm$~0.6)~$\times 10^{18}$~eV. At (11~$\pm$~1.6)~$\times 10^{18}$~eV, the elongation rate further increases to 16~$\pm$3~g~cm$^{-2}$/decade before reaching 42~$\pm$9~g~cm$^{-2}$/decade above (31~$\pm$~5)~$\times 10^{18}$~eV. Using this three-break model, the assumption of a constant elongation rate is statistically rejected with a significance greater than $4\sigma$. The trends depend very weakly on hadronic interaction models, thus providing strong constraints on possible acceleration and propagation scenarios once the mass information is used in combination with the UHECR flux~\cite{AUGER_combFit}. 
 
\subsection{Hadronic interactions}

 \begin{figure}
     \centering
      \begin{tabular}{cc}
        \includegraphics[width=0.47\textwidth]{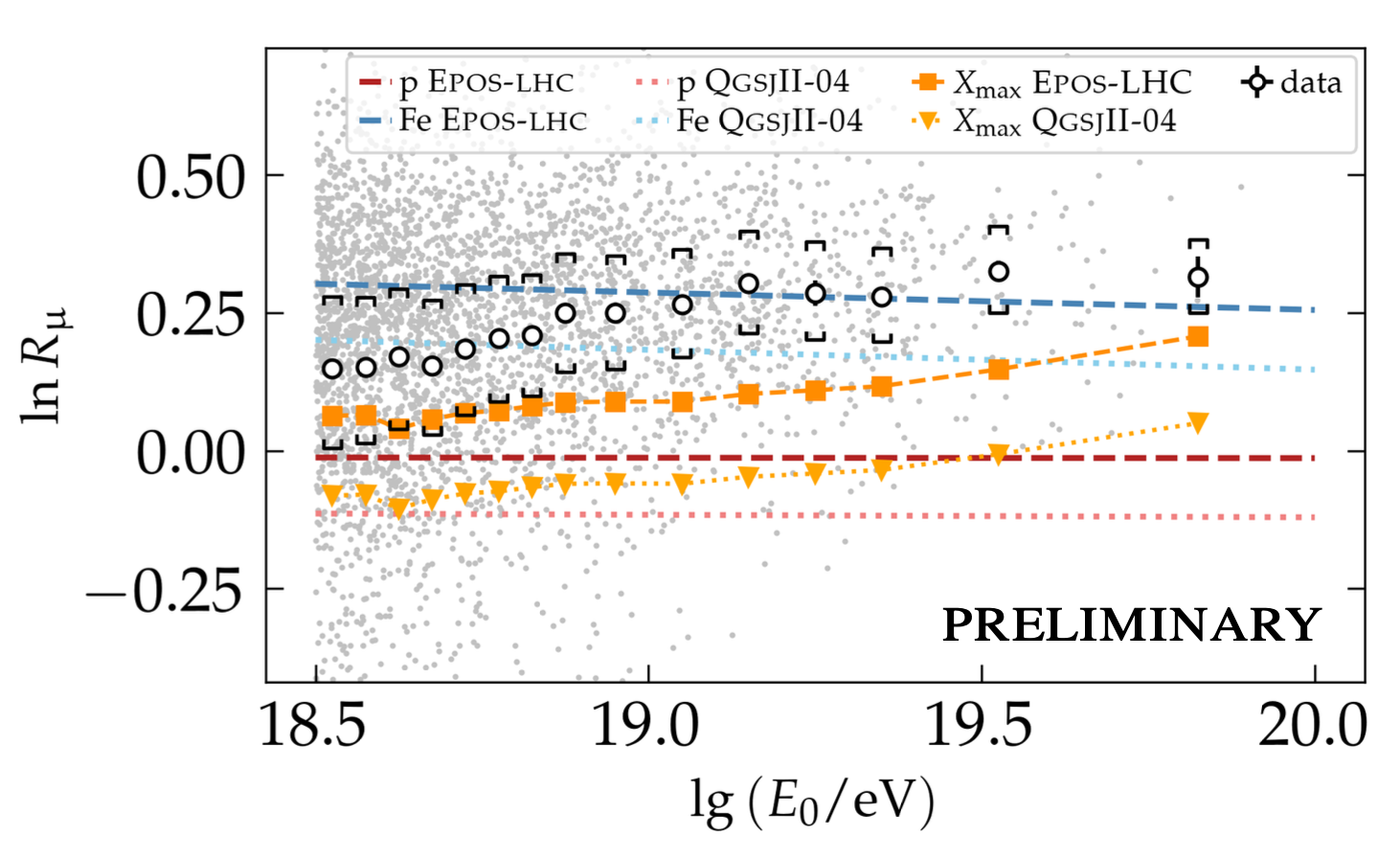} &
        \includegraphics[width=0.43\textwidth]{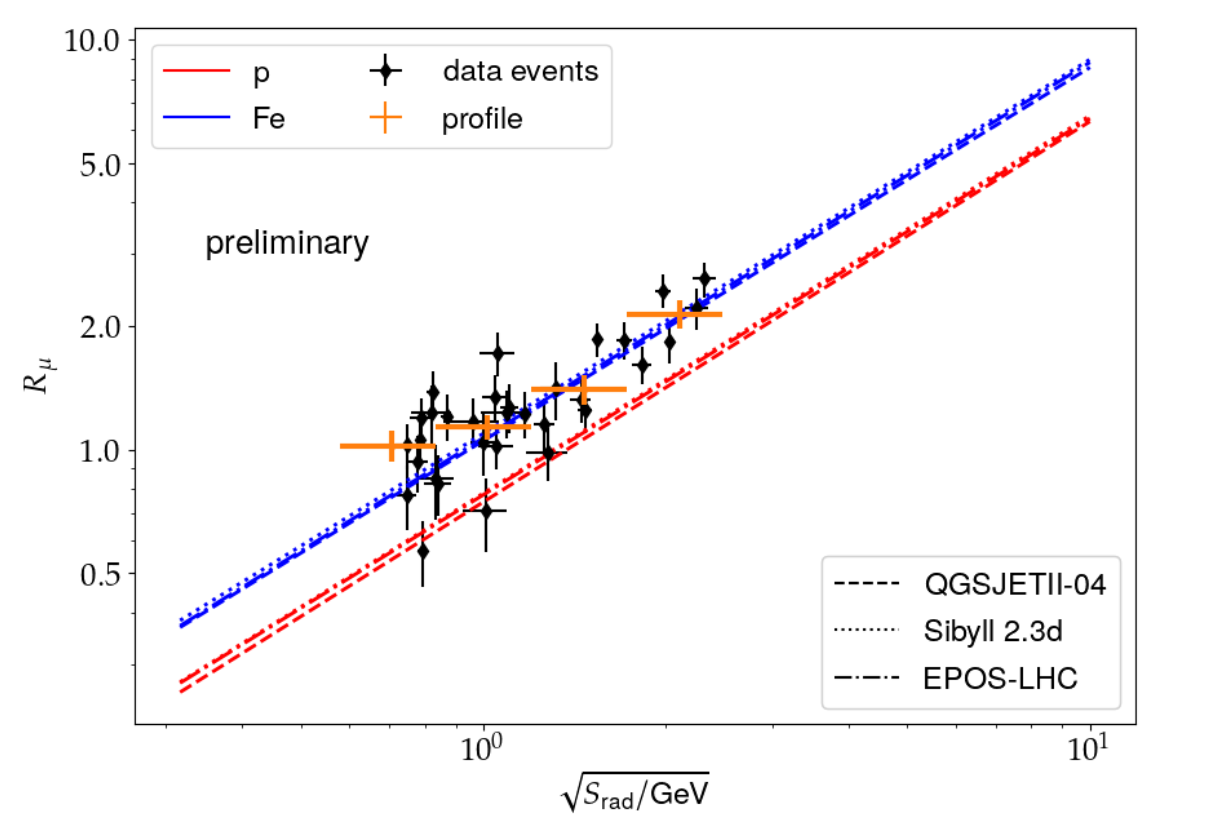}
     \end{tabular}
     \caption{Left: logarithmic muons' number obtained through combined FD and SD measurements for energies higher than $3 \times 10^{18}$~eV. The blue and red lines indicate the expected values for primaries from protons and irons simulated using the hadronic interaction models Epos-LHC and QGSjetII-04. Right: muon content as a function of RD energy estimator. Expectations from different hadronic interaction models are shown with dashed lines. Measured data are shown in black, while a profile of the data is in orange.}
     \label{fig:muon_excess}
 \end{figure}

Interpreting experimental observations in terms of the primary composition is susceptible to significant systematic uncertainties, primarily arising from the limited understanding of hadronic interactions at extremely high energies. In our efforts to mitigate this uncertainty, we conducted measurements of the attenuation length of air showers induced by protons, which are the primary contributors to the tail end of the $\textrm{X}_{\textrm{max}}$ distribution and gives the possibility to study the proton-air cross section~\cite{ICRC_olena}. Furthermore, the measurement of the muonic component at ground level proves to be highly sensitive to the intricate details of hadronic interactions throughout various stages of the cascade. In this respect, data collected by the Auger Observatory show that the number of muons predicted by the models does not agree with measurements showing an excess of muons. This feature, known as \emph{muon puzzle}, has been previously reported by different measurements~\cite{ICRC_whisp}. The left panel of  figure~\ref{fig:muon_excess} shows the logarithmic muons number obtained through combined FD and SD measurements for energies higher than $3 \times 10^{18}$~eV~\cite{ICRC_stadelmaier}. The blue and red lines indicate the expected values for protons and irons showers simulated using the hadronic interaction models Epos-LHC~\cite{eposlhc} and QGSjetII-04~\cite{qgsjet}. At the highest energies, we observe a higher average number of muons than predicted by the hadronic interaction models. The orange dots indicate the expected muon values corresponding to the converted Auger $\textrm{X}_{\textrm{max}}$ measurements assuming a hadronic interaction model. At all energies, the observed muon number is significantly higher than expected from $\textrm{X}_{\textrm{max}}$ measurements. A plot of the same type is shown in the right panel of figure~\ref{fig:muon_excess}. Here the number of muons obtained using showers simultaneously measured by the RD and SD~\cite{ICRC_gottowik} is shown as a function of the RD energy estimator $\sqrt{\textrm{S}_{\textrm{rad}}}$. The muon number is compatible with the iron simulations in a region where Auger mass measurements report a mixed composition of proton and nitrogen. On the other hand, the relative fluctuations in the number of muons measured by using the inclined FD hybrid events (62$^\circ < \theta  < 80^\circ$) of the Pierre Auger Observatory are in agreement with the model~\cite{Auger_MuonFlucPRL}. This excludes the first interaction as the main cause of this discrepancy and points towards a small effect accumulating over many generations or a very particular modification of the first interaction that changes the number of muons without changing the fluctuations~\cite{ICRC_trimarelli}.

\subsection{Arrival Directions}
 
\begin{figure}
     \centering
      \begin{tabular}{cc}
        \includegraphics[width=0.45\textwidth]{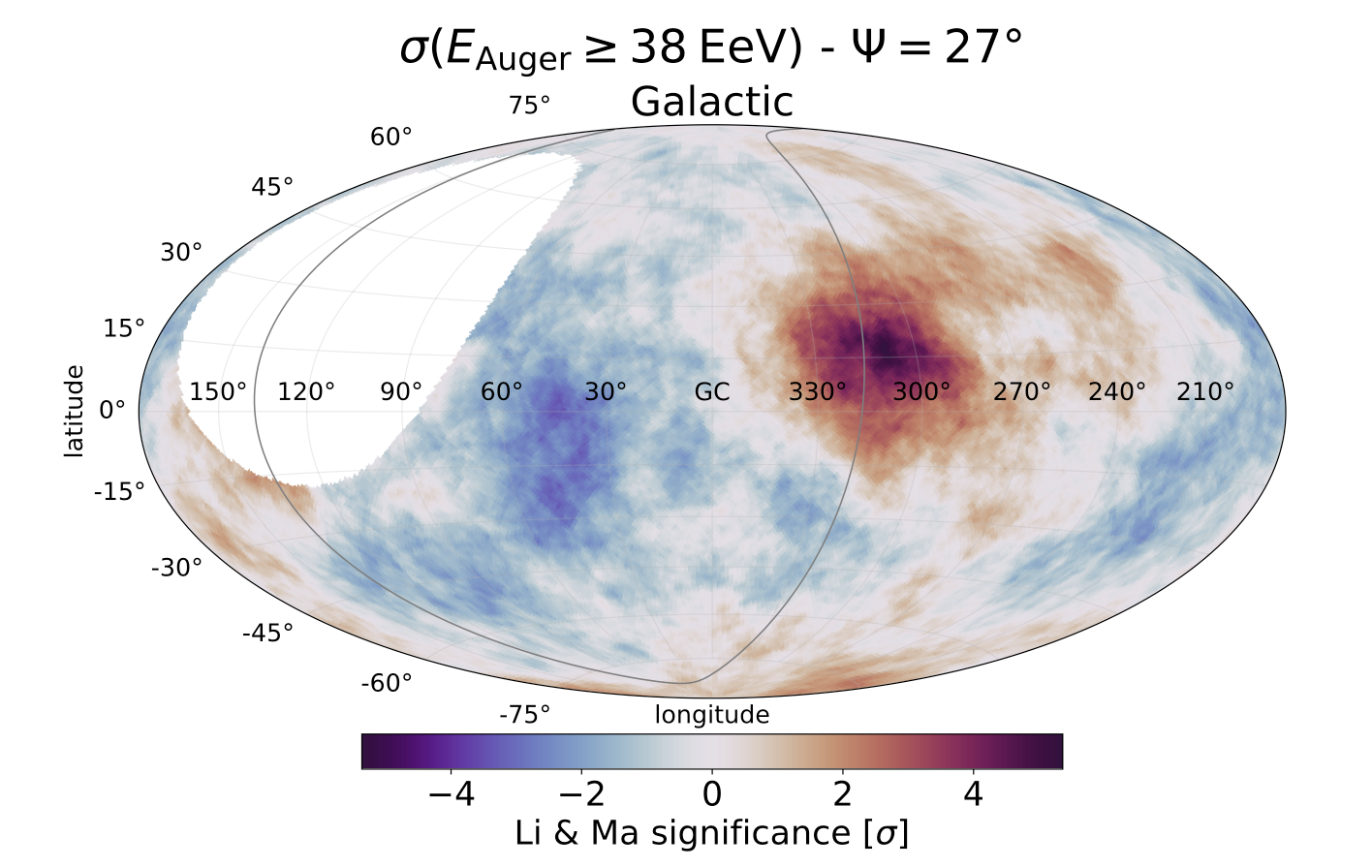} &
        \includegraphics[width=0.45\textwidth]{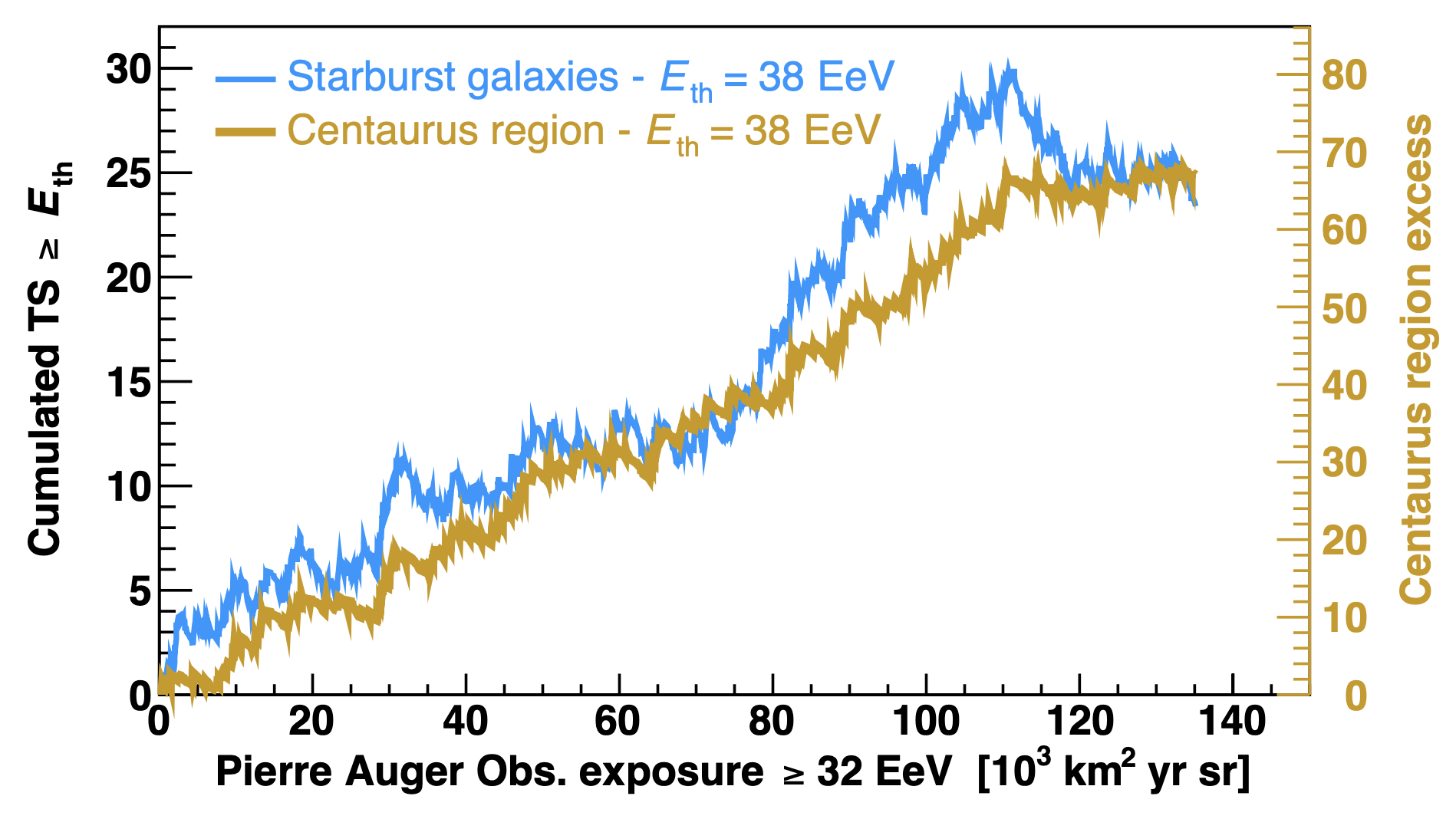}
     \end{tabular}
     \caption{Left: Li-Ma significance map within a top-hat window of
       27$^\circ$ radius with E$\ge 38 \times 10^{18}$eV in Galactic coordinates. The supergalactic plane is shown with a gray line. Right: test statistic of the starburst model and excess in the Centaurus region above the best energy threshold as a function of exposure accumulated by the Pierre Auger Observatory.}
     \label{fig:small_Scale_Ani}
 \end{figure}

A key element in understanding the origin of UHECRs is the search for anisotropies in their arrival directions. In particular, considering that magnetic deflections are proportional to the rigidity of cosmic rays E/Z and the horizon available to them decreases with increasing energy, through the study of anisotropies one can hope to backtrace the sources.
Using the events measured by the SD detector during Phase~I of the Observatory in the period January 2004 - December 2022, different analyses were carried out to search for anisotropies in the sky selecting events with energy above $3.2\times10^{19}$~eV with zenith angles lower than 80 degrees and corresponding to a total exposure of 135000~km$^2$~sr~yr and 85\% of the entire sky coverage~\cite{ICRC_golup}. The most significant results can be summarized as follows: i) the presence of an overdensity at the Centaurus region, having a post-trial $p$-value of $3.0 \times 10^{-5}$ corresponding to 4$\sigma$; ii) a likelihood test for correlation with starburst galaxies (SBG) catalog exhibiting a post-trial p-value of $6.6 \times 10^{-5}$~(3.8$\sigma$).
The angular window for the best fit in the likelihood analysis is compatible with the one found for the Centaurus region, around 25$^\circ$. This is due to the fact that also the likelihood analysis is driven by the overdensity in the Centaurus region, with two prominent SBG galaxies, NGC4945 and M83, lying relatively close to Cen~A, within the excess region.
The diminished statistical significance for the SBG case with respect to \cite{AUGER_anisAPJ}, instead, is driven by the presence of NGC253 close to the Galactic South Pole and the decrease of its flux from $\Phi_{\textrm{NGC253}} = (12. 8\pm1.2) \times
10^{-3}$km${-2}$yr$^{-1}$sr$^{-1}$ (for energies above $4\times10^{19}$eV and a top-hat window of 25$^\circ$), as reported in \cite{AUGER_anisAPJ}, to $\Phi_{\textrm{NGC253}} = (12.2 \pm 1.2) \times 10^{-3}$km$^{-2}$yr$^{-1}$sr$^{-1}$, within its statistical uncertainty. The test statistics of the SBG model, together with the excess in the Centaurus region, as a function of the exposure accumulated over time by the Observatory is shown in the right panel of figure~\ref{fig:small_Scale_Ani}.

\begin{figure}
     \centering
      \begin{tabular}{cc}
        \includegraphics[width=0.55\textwidth]{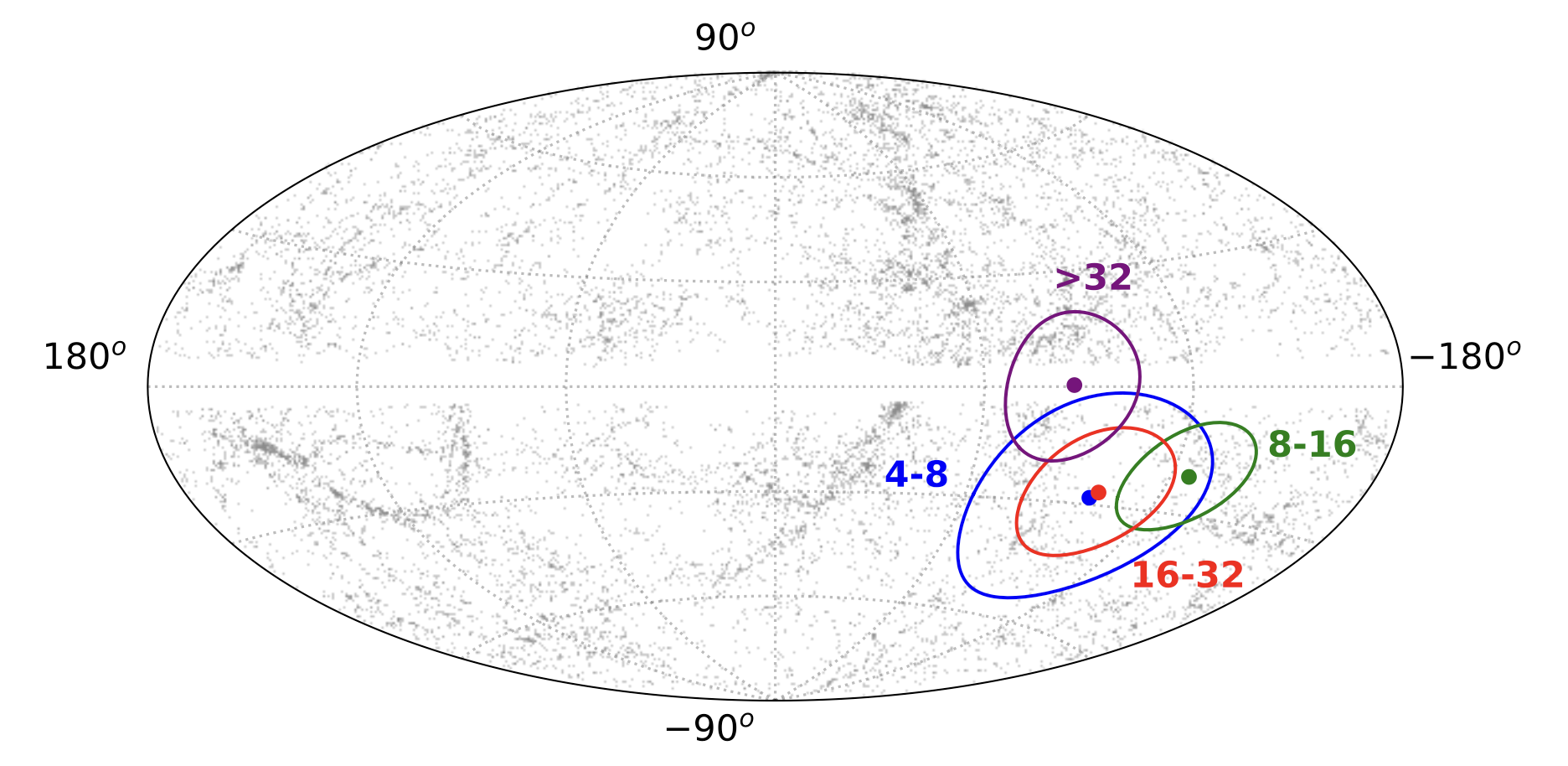} &
        \includegraphics[width=0.4\textwidth]{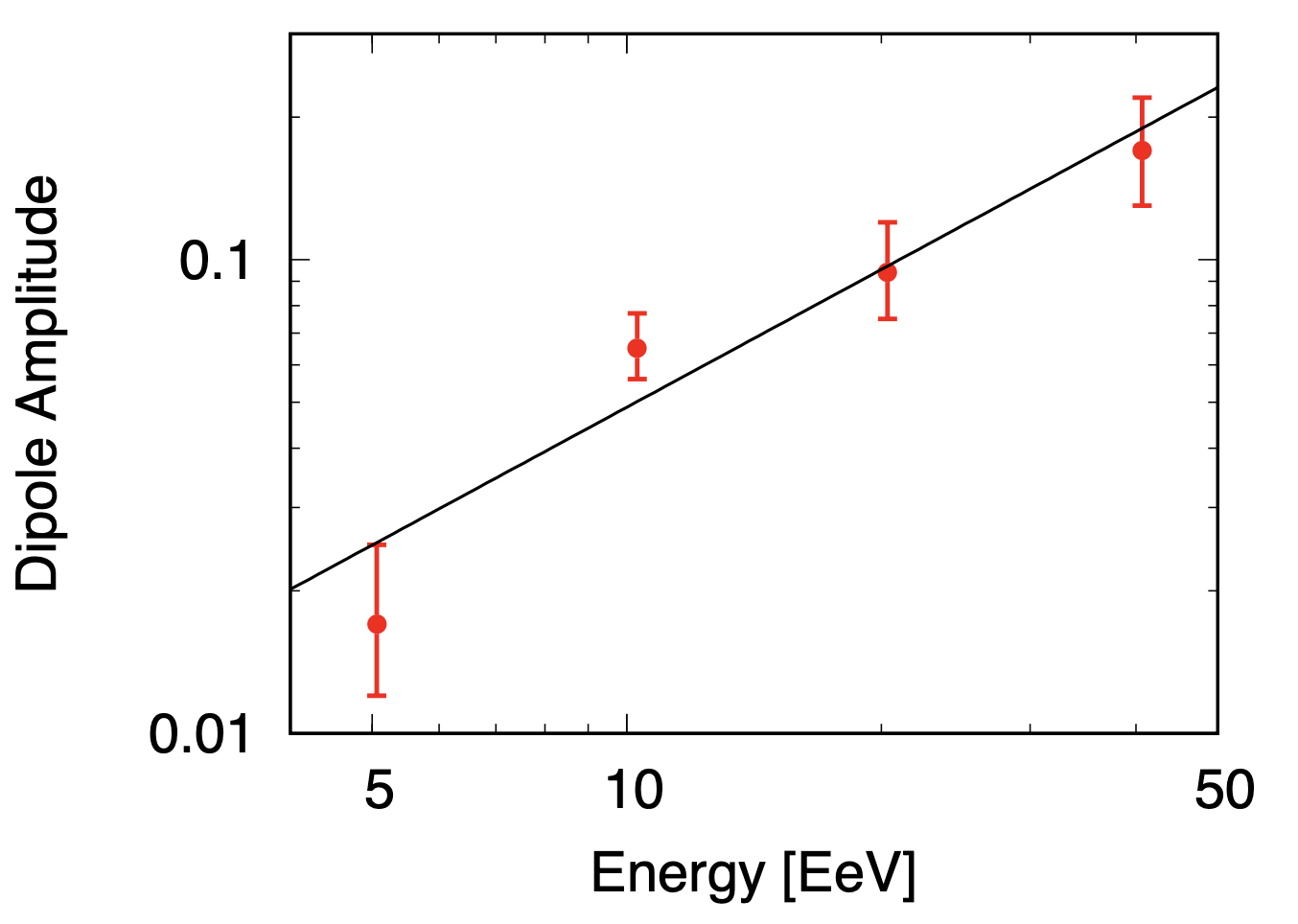}
     \end{tabular}
     \caption{Left: map with the directions of the 3D dipole for different energy bins, in Galactic coordinates. The contours of equal probability per unit solid angle, marginalized over the dipole amplitude, which contains the 68\% CL range, are also shown. The dots represent the location of the galaxies in the 2MRS catalog within 100 Mpc. Right: the evolution of the dipole amplitude with energy.}
     \label{fig:large_scale_Ani}
 \end{figure}

 Regarding the analyses performed for large angular scales, the significance of the dipolar modulation in R.A. for the cumulative energy range above $8 \times 10^{18}$~eV is now 6.9$\sigma$ and that between (8-16)$\times10^{18}$eV is 5.7$\sigma$, an increase over that shown in previous publications~\cite{AUGER_largescale1,AUGER_largescale2,ICRC_almeida}. The direction of this dipole points $113^\circ$ away from the Galactic Center, suggesting an extragalactic origin of cosmic rays above the selected energy threshold. In the left panel of figure~\ref{fig:large_scale_Ani}, the evolution of the dipole direction as a function of energy on a sky map in Galactic coordinates is shown. In the right panel of figure~\ref{fig:large_scale_Ani}, we see the growth of the dipole amplitude as the energy increases. This is maybe due to the interplay between the higher rigidity particles that are less deflected by magnetic fields and and nearby non-homogeneously located sources making a larger contribution to the flux.
 
 The promising inclusion of mass-composition estimators on an event-by-event basis with  AugerPrime and the improved mass estimators with Phase~I data, e.g.\ using DNN methods will allow more insight into the arrival direction results in the near future. At the moment, there are already some studies in the context of the Observatory that try to provide answers in this direction. In particular, hybrid events detected by the FD and with a signal in at least one SD detector can be used to separate heavy and light UHECRs using $\textrm{X}_{\textrm{max}}$ and, despite the limited FD statistics, to study whether they are distributed differently in the sky. An Anderson-Darling test has been used for comparing the cumulative $\textrm{X}_{\textrm{max}}$ distributions above $5\times 10^{18}$eV from the Galactic Plane region and those from the region outside. Already in 2021, an indication of a difference was reported~\cite{ICRC21_mayotte} and we are continuing to monitor the evolution of this indication which at the moment has a significance of about 2.5$\sigma$~\cite{ICRC_mayotte}.
 Another innovative analysis, taking into account mass-composition information from the FD, has been developed following the approach in \cite{AUGER_combFit}. We simultaneously fit the energy spectrum, the distributions of shower maxima, and the arrival directions~\cite{ICRC_bister}. The astrophysical model used incorporates uniformly distributed background sources and allows for the adjustment of the contribution of nearby candidate sources. We have also taken into account propagation effects and a rigidity-dependent magnetic field blurring, which results in an increasing level of anisotropy as energy increases.
   \begin{figure}
     \centering
     \begin{tabular}{cc}
        \includegraphics[height=0.5\textheight]{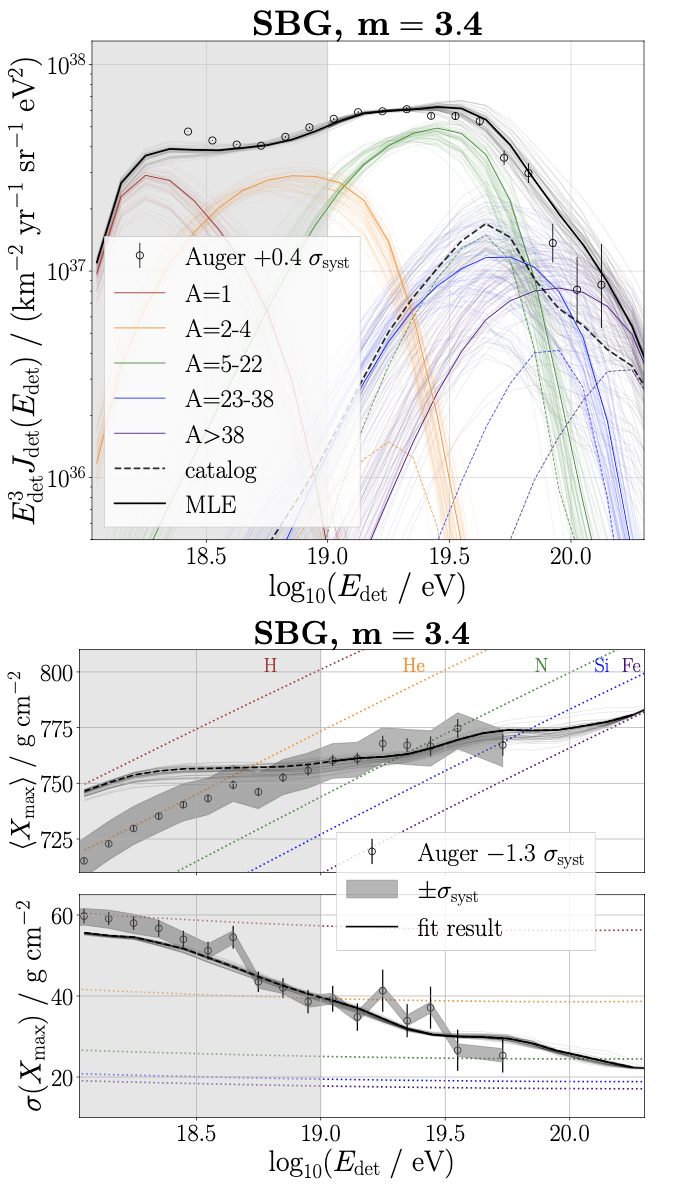} &
        \includegraphics[height=0.5\textheight]{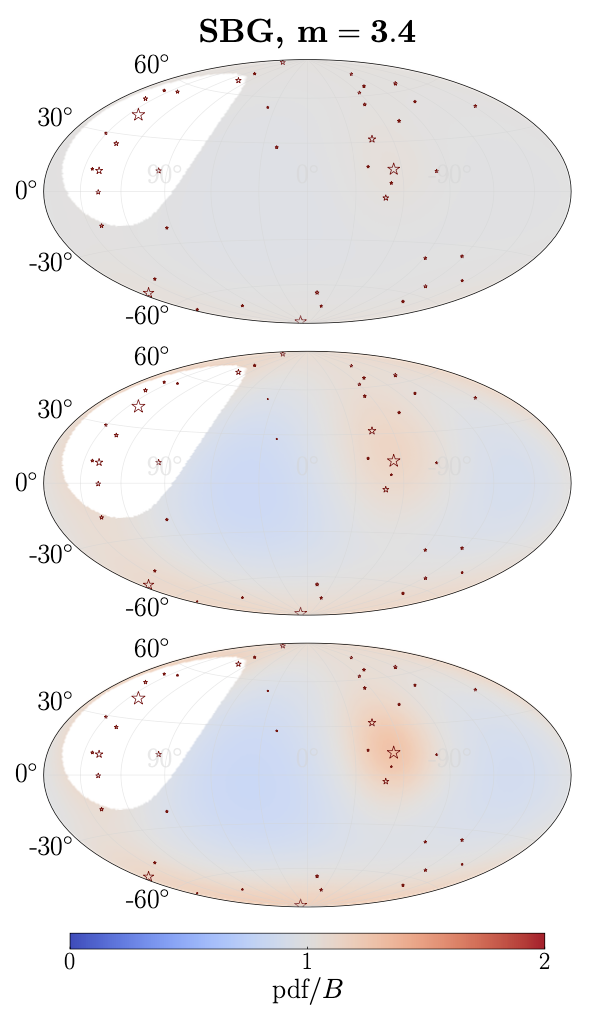}
     \end{tabular}
     \caption{Left: Fitted spectrum (upper row) and
       $\textrm{X}_{\textrm{max}}$ moments (lower row) at Earth for the SBG model with m = 3.4. The thick lines indicate the best fit, while the thin lines are drawn from the posterior distribution demonstrating the uncertainty. The measurements from the Pierre Auger Observatory are indicated with markers. Right: Modeled arrival directions for three different energy bins. The stars denote the directions of the source candidates with the size scaling with their relative flux contribution.}
     \label{fig:comb_fit_arriv_dir}
   \end{figure}
   Comparing different models with the reference one with only a homogeneous background, the best result in terms of test statistics was obtained for that incorporating the starburst galaxies, with a significance of $4.5 \sigma$~(compared to $4.0 \sigma$ in \cite{ICRC_almeida}). The best fit was obtained with a much harder spectral index ($\gamma \sim -1$) than expected from diffusive shocks acceleration, a signal fraction of about 20\% at $4\times 10^{19}$~eV and a magnetic field blurring of about 20$^\circ$ for a particle rigidity of 10 EV. The corresponding fitted spectrum and $\textrm{X}_{\textrm{max}}$ moments distributions are illustrated in the left panel of figure~\ref{fig:comb_fit_arriv_dir}. The thick lines indicate the best fit, while the thin lines are drawn from the posterior distribution demonstrating the uncertainty. Our measurements are indicated with markers. The grey area refers to the energy bins that are not fully included in the fit as described in  \cite{ICRC_bister}. 
   The right panel of figure~\ref{fig:comb_fit_arriv_dir} displays the modeled arrival directions for three different energy bins, providing a clear illustration of the important role of the Centaurus region in the fit of blurring and signal fraction. Additionally, other notable observations can be made: as energy grows, the degree of anisotropy becomes more pronounced, because of the increased impact of nearby catalogue sources due to propagation effects. A model in which the arrival directions are correlated with $\gamma$-AGNs  was also tested; we can conclude that it does not fit well with the modeling of arrival directions as a function of energy. This is mainly due to the strong contribution of the distant blazar Markarian 421. Therefore, a scenario in which the UHECRs intensity scales with the $\gamma$-ray flux for a $\gamma$-AGN source model is disfavored.

\subsection{Multimessenger Astrophysics}

\begin{figure}
     \centering
      \begin{tabular}{cc}
        \includegraphics[width=0.45\textwidth]{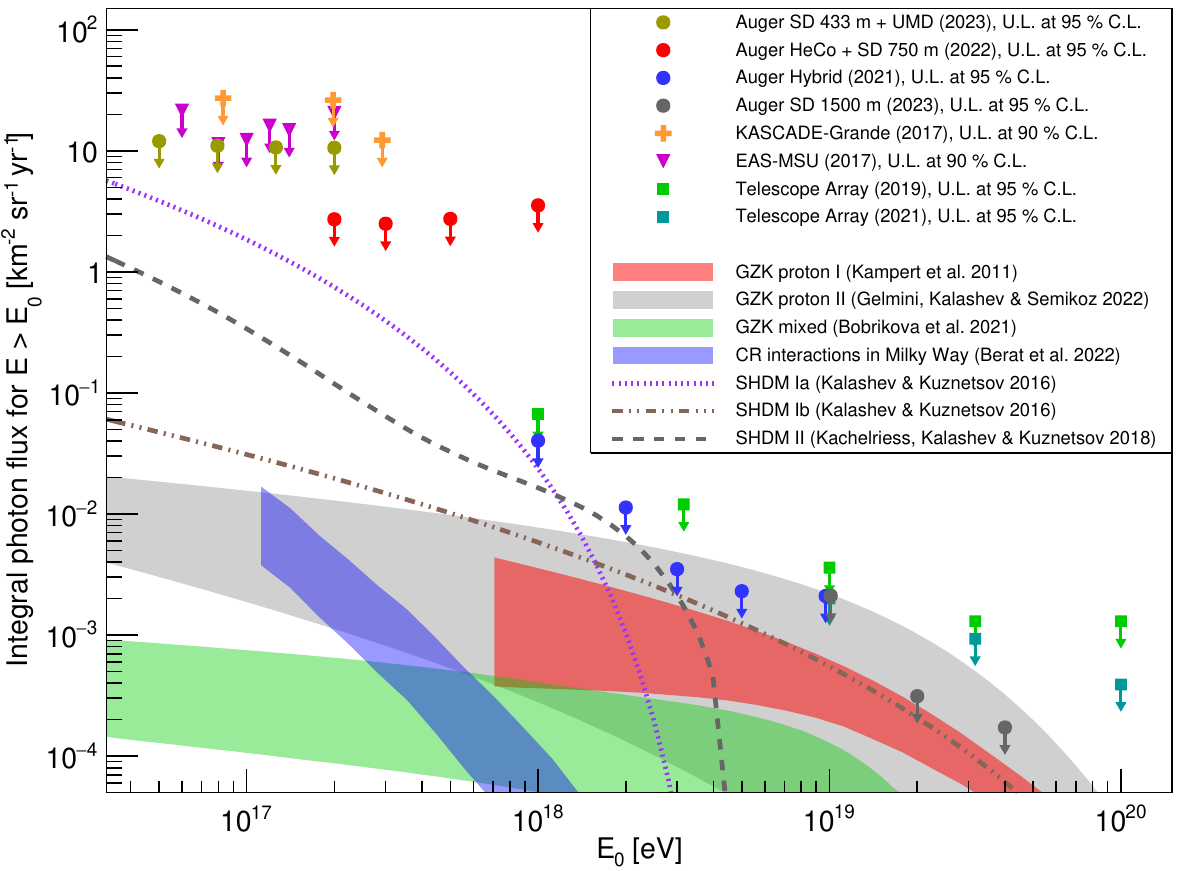} &
        \includegraphics[width=0.55\textwidth]{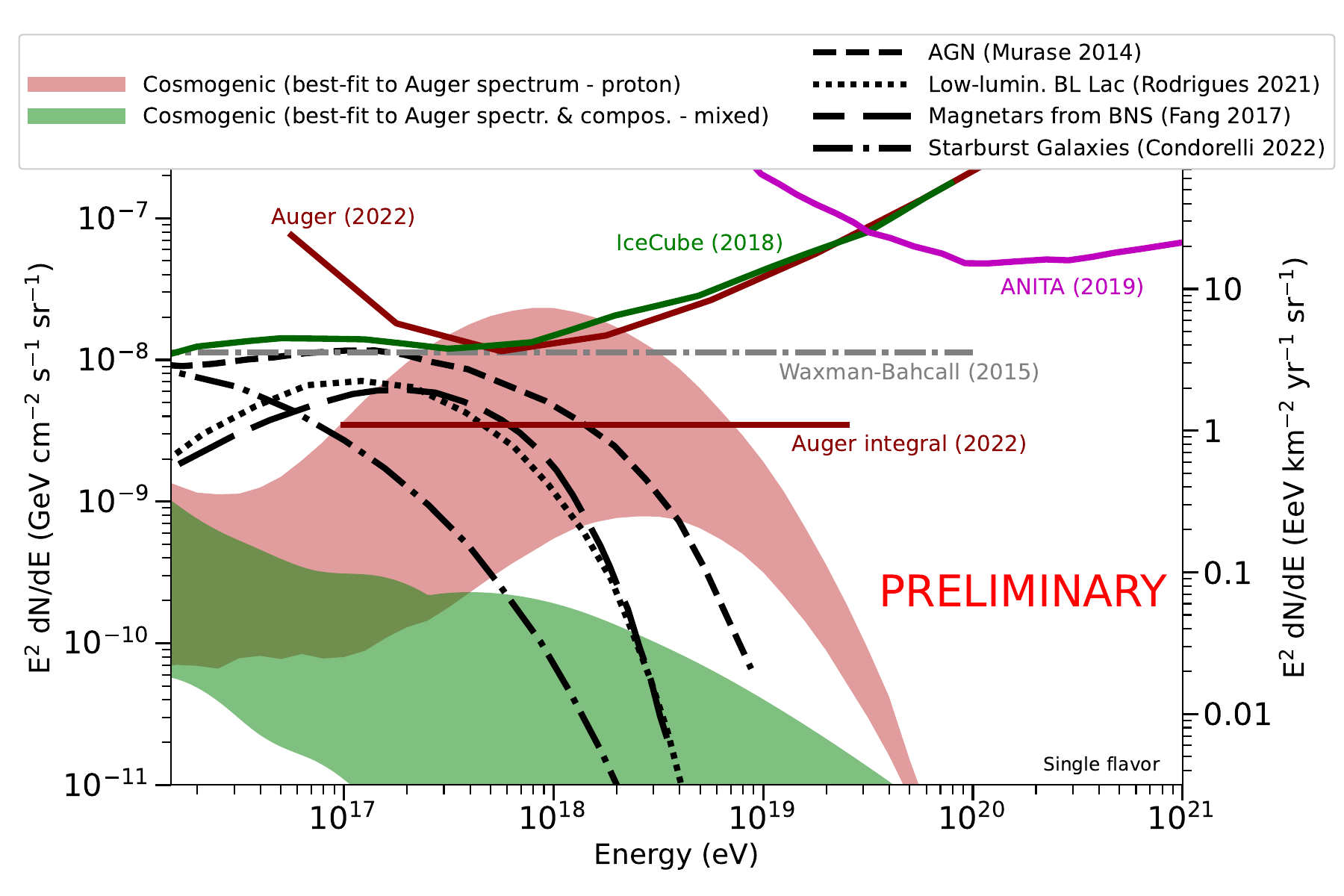}
     \end{tabular}
     \caption{Left: upper limits on the integral photon flux determined from data collected by the Pierre Auger Observatory (red, blue, and gray circles).  Right: upper limits on the diffuse flux of neutrinos, both integral (straight solid line) as well as differential (curved solid line). }
     \label{fig:augerMultimessenger}
 \end{figure}

In the context of multimessenger astronomy, the Pierre Auger Observatory offers the unique possibility to aim at the detection of high-energy photons and neutrinos above 10$^{17}$eV. Because they are not subject to magnetic fields, they propagate in a straight line through the interstellar medium and can originate directly from the sources (i.e.\ \emph{astrophysical}) or during their journey towards us (i.e.\ \emph{cosmogenic}). Both, if measured, can provide valuable information about the sources and the mechanisms of acceleration. Photons are effective at tracing the local Universe on scales up to megaparsecs (Mpc), as their interactions with the cosmic background fields limit their range. On the other hand, neutrinos could easily traverse the entire Universe without interacting. Furthermore, both photons and neutrinos can serve as valuable tools for investigating fundamental physics, offering insights into areas such as super-heavy dark matter~\cite{auger_SHDM}, potential violations of Lorentz invariance~\cite{auger_liv}, GW follow-up, and in general physics beyond the standard model~\cite{ICRC_devito, ICRC_yue}. 
The analyses for the identification of primary photons are based on the fact that the development of photon-induced showers in the atmosphere is dominated by electromagnetic interactions, which produce events with a deeper X$_{\textrm{max}}$ and a smaller muon component than those induced by protons and nuclei. Using the Phase~I measurements from the 1500~m array over a data-taking period of about 16 years (01/01/2004-30/06/2020), it was possible to search for events compatible with photons of energies above $10^{19}$~eV. The analysis is based on a Fisher discriminant using the variables L$_{LDF}$ and $\Delta$ linking the total measured signals at individual SD stations and the measured rise times of the signals with respect to a data reference, which describe the average of all SD data~\cite{augerPhotons, ICRC_niechciol}. Despite having found 16 events that fulfill the criteria to be considered potential photon candidates, this number is consistent with the expected background. Consequently, an upper limit has been placed on the cumulative flux of primary photons. This limit, along with upper limits derived from other experiments and photon fluxes predicted based on various hypotheses and theoretical scenarios, is shown in the left panel of the figure~\ref{fig:augerMultimessenger}. In particular, the upper limits established by the Pierre Auger Observatory are the most stringent to date, covering a wide energy range from $5 \times 10^{16}$~eV to the highest energies. Additionally, the figure displays the limits obtained by the Pierre Auger Observatory at lower energies using the FD measurements (E$\ge 2 \times 10^{17}$~eV) and preliminary results from an extension to even lower energies, above $5 \times 10^{16}$~eV, using the dense the data measured simultaneously by the 433~m SD array and the UMD~\cite{ICRC_gonzalez}. These current limits have significantly constrained certain "exotic" top-down models and predictions from some cosmogenic models.
Searches for UHE neutrinos have been performed by exploiting those inclined showers characterized by a prominent electromagnetic component.  In fact, for protons and nuclei at large zenith angles, the atmosphere fully absorbs the electromagnetic component, resulting in only muons reaching the ground. On the contrary, neutrinos, having a smaller interaction cross-section, can also initiate showers near the ground, allowing the electromagnetic component to be detected even for significant inclined events. Using the data from the 1500 m SD array collected between 1 January 2004 and
31 December 2021, upper limits on the diffuse flux of UHE neutrinos have been obtained~\cite{Auger_nu_JCAP}.  Figure~\ref{fig:augerMultimessenger}~(right) displays the upper limits alongside those determined by IceCube and ANITA experiments, as well as the expected neutrino fluxes based on various theoretical assumptions and scenarios. Using the expected number of cosmogenic neutrinos from simulations~\cite{simprop}, it is possible to constrain different UHECR models by considering the combinations of spectral parameters and mass composition at their sources, as well as parameters related to the source distribution~\cite{ICRC_petrucci}. Notably, some cosmogenic models characterized by a composition consisting of pure protons and exhibiting a strong source evolution with redshift have been constrained, and some of them were even ruled out, due to the absence of neutrino observations so far.

 \section{Prospects}
  UHECRs remain enigmatic despite more than 60 years having passed since the first detection by Linsley and their origin remains largely unknown. The Pierre Auger Observatory has made significant progress, collecting extensive data, though the cosmic ray source remains elusive. The enhanced Pierre Auger Observatory will enter a new era of data analysis by exploiting the multi-hybrid measurements it will offer. The different analyses, here presented, will harness the capabilities of both the existing and newly integrated detector components, along with the upgraded station electronics and broader dynamic range.
  The simultaneous analysis of the signals from the different detectors will make it possible in the near future to obtain an estimate of the primary mass on an event-by-event basis by separating the muon and electromagnetic components. Machine learning (ML) techniques have already demonstrated their potential by extracting valuable information on $X_{\textrm{max}}$ from WCD data alone, and by combining the Phase~I data set with the AugerPrime data statistics for mass composition studies will more than double by the end of Phase~II. In addition, the promising inclusion of mass-composition estimators on an event-by-event basis with AugerPrime and the improved mass estimators with Phase~I data will allow more insight into the arrival direction results in the near future. 
  The Observatory in the coming years will also offer the possibility of expanding the scope of its investigations in the field of cosmo-geophysics~\cite{ICRC_mussa, ICRC_colalillo} and with the Auger Open Data Portal will offer a good overview of the detectors and the achievements of the Collaboration, as well as a direct invitation to the general public to use the released data~\cite{ICRC_ghia, ICRC_sarmento}.

\clearpage

\section*{The Pierre Auger Collaboration}
\small

\begin{wrapfigure}[8]{l}{0.11\linewidth}
\vspace{-5mm}
\includegraphics[width=0.98\linewidth]{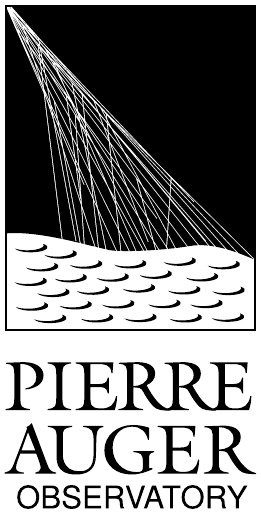}
\end{wrapfigure}
\begin{sloppypar}\noindent
A.~Abdul Halim$^{13}$,
P.~Abreu$^{72}$,
M.~Aglietta$^{54,52}$,
I.~Allekotte$^{1}$,
K.~Almeida Cheminant$^{70}$,
A.~Almela$^{7,12}$,
R.~Aloisio$^{45,46}$,
J.~Alvarez-Mu\~niz$^{79}$,
J.~Ammerman Yebra$^{79}$,
G.A.~Anastasi$^{54,52}$
% ...

\end{sloppypar}

\begin{center}
\rule{0.1\columnwidth}{0.5pt}
\raisebox{-0.4ex}{\scriptsize$\bullet$}
\rule{0.1\columnwidth}{0.5pt}
\end{center}

\vspace{-1ex}
\footnotesize
\begin{description}[labelsep=0.2em,align=right,labelwidth=0.7em,labelindent=0em,leftmargin=2em,noitemsep]
\item[$^{1}$] Centro At\'omico Bariloche and Instituto Balseiro (CNEA-UNCuyo-CONICET), San Carlos de Bariloche, Argentina
\item[$^{2}$] Departamento de F\'\i{}sica and Departamento de Ciencias de la Atm\'osfera y los Oc\'eanos, FCEyN, Universidad de Buenos Aires and CONICET, Buenos Aires, Argentina
\item[$^{3}$] IFLP, Universidad Nacional de La Plata and CONICET, La Plata, Argentina
\item[$^{4}$] Instituto de Astronom\'\i{}a y F\'\i{}sica del Espacio (IAFE, CONICET-UBA), Buenos Aires, Argentina
\item[$^{5}$] Instituto de F\'\i{}sica de Rosario (IFIR) -- CONICET/U.N.R.\ and Facultad de Ciencias Bioqu\'\i{}micas y Farmac\'euticas U.N.R., Rosario, Argentina
\item[$^{6}$] Instituto de Tecnolog\'\i{}as en Detecci\'on y Astropart\'\i{}culas (CNEA, CONICET, UNSAM), and Universidad Tecnol\'ogica Nacional -- Facultad Regional Mendoza (CONICET/CNEA), Mendoza, Argentina
\item[$^{7}$] Instituto de Tecnolog\'\i{}as en Detecci\'on y Astropart\'\i{}culas (CNEA, CONICET, UNSAM), Buenos Aires, Argentina
\item[$^{8}$] International Center of Advanced Studies and Instituto de Ciencias F\'\i{}sicas, ECyT-UNSAM and CONICET, Campus Miguelete -- San Mart\'\i{}n, Buenos Aires, Argentina
\item[$^{9}$] Laboratorio Atm\'osfera -- Departamento de Investigaciones en L\'aseres y sus Aplicaciones -- UNIDEF (CITEDEF-CONICET), Argentina
\item[$^{10}$] Observatorio Pierre Auger, Malarg\"ue, Argentina
\item[$^{11}$] Observatorio Pierre Auger and Comisi\'on Nacional de Energ\'\i{}a At\'omica, Malarg\"ue, Argentina
\item[$^{12}$] Universidad Tecnol\'ogica Nacional -- Facultad Regional Buenos Aires, Buenos Aires, Argentina
\item[$^{13}$] University of Adelaide, Adelaide, S.A., Australia
\item[$^{14}$] Universit\'e Libre de Bruxelles (ULB), Brussels, Belgium
\item[$^{15}$] Vrije Universiteit Brussels, Brussels, Belgium
\item[$^{16}$] Centro Federal de Educa\c{c}\~ao Tecnol\'ogica Celso Suckow da Fonseca, Petropolis, Brazil
\item[$^{17}$] Instituto Federal de Educa\c{c}\~ao, Ci\^encia e Tecnologia do Rio de Janeiro (IFRJ), Brazil
\item[$^{18}$] Universidade de S\~ao Paulo, Escola de Engenharia de Lorena, Lorena, SP, Brazil
\item[$^{19}$] Universidade de S\~ao Paulo, Instituto de F\'\i{}sica de S\~ao Carlos, S\~ao Carlos, SP, Brazil
\item[$^{20}$] Universidade de S\~ao Paulo, Instituto de F\'\i{}sica, S\~ao Paulo, SP, Brazil
\item[$^{21}$] Universidade Estadual de Campinas, IFGW, Campinas, SP, Brazil
\item[$^{22}$] Universidade Estadual de Feira de Santana, Feira de Santana, Brazil
\item[$^{23}$] Universidade Federal de Campina Grande, Centro de Ciencias e Tecnologia, Campina Grande, Brazil
\item[$^{24}$] Universidade Federal do ABC, Santo Andr\'e, SP, Brazil
\item[$^{25}$] Universidade Federal do Paran\'a, Setor Palotina, Palotina, Brazil
\item[$^{26}$] Universidade Federal do Rio de Janeiro, Instituto de F\'\i{}sica, Rio de Janeiro, RJ, Brazil
\item[$^{27}$] Universidade Federal do Rio de Janeiro (UFRJ), Observat\'orio do Valongo, Rio de Janeiro, RJ, Brazil
\item[$^{28}$] Universidade Federal Fluminense, EEIMVR, Volta Redonda, RJ, Brazil
\item[$^{29}$] Universidad de Medell\'\i{}n, Medell\'\i{}n, Colombia
\item[$^{30}$] Universidad Industrial de Santander, Bucaramanga, Colombia
\item[$^{31}$] Charles University, Faculty of Mathematics and Physics, Institute of Particle and Nuclear Physics, Prague, Czech Republic
\item[$^{32}$] Institute of Physics of the Czech Academy of Sciences, Prague, Czech Republic
\item[$^{33}$] Palacky University, Olomouc, Czech Republic
\item[$^{34}$] CNRS/IN2P3, IJCLab, Universit\'e Paris-Saclay, Orsay, France
\item[$^{35}$] Laboratoire de Physique Nucl\'eaire et de Hautes Energies (LPNHE), Sorbonne Universit\'e, Universit\'e de Paris, CNRS-IN2P3, Paris, France
\item[$^{36}$] Univ.\ Grenoble Alpes, CNRS, Grenoble Institute of Engineering Univ.\ Grenoble Alpes, LPSC-IN2P3, 38000 Grenoble, France
\item[$^{37}$] Universit\'e Paris-Saclay, CNRS/IN2P3, IJCLab, Orsay, France
\item[$^{38}$] Bergische Universit\"at Wuppertal, Department of Physics, Wuppertal, Germany
\item[$^{39}$] Karlsruhe Institute of Technology (KIT), Institute for Experimental Particle Physics, Karlsruhe, Germany
\item[$^{40}$] Karlsruhe Institute of Technology (KIT), Institut f\"ur Prozessdatenverarbeitung und Elektronik, Karlsruhe, Germany
\item[$^{41}$] Karlsruhe Institute of Technology (KIT), Institute for Astroparticle Physics, Karlsruhe, Germany
\item[$^{42}$] RWTH Aachen University, III.\ Physikalisches Institut A, Aachen, Germany
\item[$^{43}$] Universit\"at Hamburg, II.\ Institut f\"ur Theoretische Physik, Hamburg, Germany
\item[$^{44}$] Universit\"at Siegen, Department Physik -- Experimentelle Teilchenphysik, Siegen, Germany
\item[$^{45}$] Gran Sasso Science Institute, L'Aquila, Italy
\item[$^{46}$] INFN Laboratori Nazionali del Gran Sasso, Assergi (L'Aquila), Italy
\item[$^{47}$] INFN, Sezione di Catania, Catania, Italy
\item[$^{48}$] INFN, Sezione di Lecce, Lecce, Italy
\item[$^{49}$] INFN, Sezione di Milano, Milano, Italy
\item[$^{50}$] INFN, Sezione di Napoli, Napoli, Italy
\item[$^{51}$] INFN, Sezione di Roma ``Tor Vergata'', Roma, Italy
\item[$^{52}$] INFN, Sezione di Torino, Torino, Italy
\item[$^{53}$] Istituto di Astrofisica Spaziale e Fisica Cosmica di Palermo (INAF), Palermo, Italy
\item[$^{54}$] Osservatorio Astrofisico di Torino (INAF), Torino, Italy
\item[$^{55}$] Politecnico di Milano, Dipartimento di Scienze e Tecnologie Aerospaziali , Milano, Italy
\item[$^{56}$] Universit\`a del Salento, Dipartimento di Matematica e Fisica ``E.\ De Giorgi'', Lecce, Italy
\item[$^{57}$] Universit\`a dell'Aquila, Dipartimento di Scienze Fisiche e Chimiche, L'Aquila, Italy
\item[$^{58}$] Universit\`a di Catania, Dipartimento di Fisica e Astronomia ``Ettore Majorana``, Catania, Italy
\item[$^{59}$] Universit\`a di Milano, Dipartimento di Fisica, Milano, Italy
\item[$^{60}$] Universit\`a di Napoli ``Federico II'', Dipartimento di Fisica ``Ettore Pancini'', Napoli, Italy
\item[$^{61}$] Universit\`a di Palermo, Dipartimento di Fisica e Chimica ''E.\ Segr\`e'', Palermo, Italy
\item[$^{62}$] Universit\`a di Roma ``Tor Vergata'', Dipartimento di Fisica, Roma, Italy
\item[$^{63}$] Universit\`a Torino, Dipartimento di Fisica, Torino, Italy
\item[$^{64}$] Benem\'erita Universidad Aut\'onoma de Puebla, Puebla, M\'exico
\item[$^{65}$] Unidad Profesional Interdisciplinaria en Ingenier\'\i{}a y Tecnolog\'\i{}as Avanzadas del Instituto Polit\'ecnico Nacional (UPIITA-IPN), M\'exico, D.F., M\'exico
\item[$^{66}$] Universidad Aut\'onoma de Chiapas, Tuxtla Guti\'errez, Chiapas, M\'exico
\item[$^{67}$] Universidad Michoacana de San Nicol\'as de Hidalgo, Morelia, Michoac\'an, M\'exico
\item[$^{68}$] Universidad Nacional Aut\'onoma de M\'exico, M\'exico, D.F., M\'exico
\item[$^{69}$] Universidad Nacional de San Agustin de Arequipa, Facultad de Ciencias Naturales y Formales, Arequipa, Peru
\item[$^{70}$] Institute of Nuclear Physics PAN, Krakow, Poland
\item[$^{71}$] University of \L{}\'od\'z, Faculty of High-Energy Astrophysics,\L{}\'od\'z, Poland
\item[$^{72}$] Laborat\'orio de Instrumenta\c{c}\~ao e F\'\i{}sica Experimental de Part\'\i{}culas -- LIP and Instituto Superior T\'ecnico -- IST, Universidade de Lisboa -- UL, Lisboa, Portugal
\item[$^{73}$] ``Horia Hulubei'' National Institute for Physics and Nuclear Engineering, Bucharest-Magurele, Romania
\item[$^{74}$] Institute of Space Science, Bucharest-Magurele, Romania
\item[$^{75}$] University Politehnica of Bucharest, Bucharest, Romania
\item[$^{76}$] Center for Astrophysics and Cosmology (CAC), University of Nova Gorica, Nova Gorica, Slovenia
\item[$^{77}$] Experimental Particle Physics Department, J.\ Stefan Institute, Ljubljana, Slovenia
\item[$^{78}$] Universidad de Granada and C.A.F.P.E., Granada, Spain
\item[$^{79}$] Instituto Galego de F\'\i{}sica de Altas Enerx\'\i{}as (IGFAE), Universidade de Santiago de Compostela, Santiago de Compostela, Spain
\item[$^{80}$] IMAPP, Radboud University Nijmegen, Nijmegen, The Netherlands
\item[$^{81}$] Nationaal Instituut voor Kernfysica en Hoge Energie Fysica (NIKHEF), Science Park, Amsterdam, The Netherlands
\item[$^{82}$] Stichting Astronomisch Onderzoek in Nederland (ASTRON), Dwingeloo, The Netherlands
\item[$^{83}$] Universiteit van Amsterdam, Faculty of Science, Amsterdam, The Netherlands
\item[$^{84}$] Case Western Reserve University, Cleveland, OH, USA
\item[$^{85}$] Colorado School of Mines, Golden, CO, USA
\item[$^{86}$] Department of Physics and Astronomy, Lehman College, City University of New York, Bronx, NY, USA
\item[$^{87}$] Michigan Technological University, Houghton, MI, USA
\item[$^{88}$] New York University, New York, NY, USA
\item[$^{89}$] University of Chicago, Enrico Fermi Institute, Chicago, IL, USA
\item[$^{90}$] University of Delaware, Department of Physics and Astronomy, Bartol Research Institute, Newark, DE, USA
\item[$^{91}$] University of Wisconsin-Madison, Department of Physics and WIPAC, Madison, WI, USA
\item[] -----
\item[$^{a}$] Louisiana State University, Baton Rouge, LA, USA
\item[$^{b}$] also at University of Bucharest, Physics Department, Bucharest, Romania
\item[$^{c}$] School of Physics and Astronomy, University of Leeds, Leeds, United Kingdom
\item[$^{d}$] now at Agenzia Spaziale Italiana (ASI).\ Via del Politecnico 00133, Roma, Italy
\item[$^{e}$] Fermi National Accelerator Laboratory, Fermilab, Batavia, IL, USA
\item[$^{f}$] now at Graduate School of Science, Osaka Metropolitan University, Osaka, Japan
\item[$^{g}$] now at ECAP, Erlangen, Germany
\item[$^{h}$] Max-Planck-Institut f\"ur Radioastronomie, Bonn, Germany
\item[$^{i}$] also at Kapteyn Institute, University of Groningen, Groningen, The Netherlands
\item[$^{j}$] Colorado State University, Fort Collins, CO, USA
\item[$^{k}$] Pennsylvania State University, University Park, PA, USA
\end{description}

\vspace{-1ex}
\footnotesize
\section*{Acknowledgments}

\begin{sloppypar}
The successful installation, commissioning, and operation of the Pierre
Auger Observatory would not have been possible without the strong
commitment and effort from the technical and administrative staff in
Malarg\"ue. We are very grateful to the following agencies and
organizations for financial support:
\end{sloppypar}

\begin{sloppypar}
Argentina -- Comisi\'on Nacional de Energ\'\i{}a At\'omica; Agencia Nacional de
Promoci\'on Cient\'\i{}fica y Tecnol\'ogica (ANPCyT); Consejo Nacional de
Investigaciones Cient\'\i{}ficas y T\'ecnicas (CONICET); Gobierno de la
Provincia de Mendoza; Municipalidad de Malarg\"ue; NDM Holdings and Valle
Las Le\~nas; in gratitude for their continuing cooperation over land
access; Australia -- the Australian Research Council; Belgium -- Fonds
de la Recherche Scientifique (FNRS); Research Foundation Flanders (FWO);
Brazil -- Conselho Nacional de Desenvolvimento Cient\'\i{}fico e Tecnol\'ogico
(CNPq); Financiadora de Estudos e Projetos (FINEP); Funda\c{c}\~ao de Amparo \`a
Pesquisa do Estado de Rio de Janeiro (FAPERJ); S\~ao Paulo Research
Foundation (FAPESP) Grants No.~2019/10151-2, No.~2010/07359-6 and
No.~1999/05404-3; Minist\'erio da Ci\^encia, Tecnologia, Inova\c{c}\~oes e
Comunica\c{c}\~oes (MCTIC); Czech Republic -- Grant No.~MSMT CR LTT18004,
LM2015038, LM2018102, CZ.02.1.01/0.0/0.0/16{\textunderscore}013/0001402,
CZ.02.1.01/0.0/0.0/18{\textunderscore}046/0016010 and
CZ.02.1.01/0.0/0.0/17{\textunderscore}049/0008422; France -- Centre de Calcul
IN2P3/CNRS; Centre National de la Recherche Scientifique (CNRS); Conseil
R\'egional Ile-de-France; D\'epartement Physique Nucl\'eaire et Corpusculaire
(PNC-IN2P3/CNRS); D\'epartement Sciences de l'Univers (SDU-INSU/CNRS);
Institut Lagrange de Paris (ILP) Grant No.~LABEX ANR-10-LABX-63 within
the Investissements d'Avenir Programme Grant No.~ANR-11-IDEX-0004-02;
Germany -- Bundesministerium f\"ur Bildung und Forschung (BMBF); Deutsche
Forschungsgemeinschaft (DFG); Finanzministerium Baden-W\"urttemberg;
Helmholtz Alliance for Astroparticle Physics (HAP);
Helmholtz-Gemeinschaft Deutscher Forschungszentren (HGF); Ministerium
f\"ur Kultur und Wissenschaft des Landes Nordrhein-Westfalen; Ministerium
f\"ur Wissenschaft, Forschung und Kunst des Landes Baden-W\"urttemberg;
Italy -- Istituto Nazionale di Fisica Nucleare (INFN); Istituto
Nazionale di Astrofisica (INAF); Ministero dell'Istruzione,
dell'Universit\'a e della Ricerca (MIUR); CETEMPS Center of Excellence;
Ministero degli Affari Esteri (MAE), ICSC Centro Nazionale di Ricerca in
High Performance Computing, Big Data and Quantum Computing, funded by
European Union NextGenerationEU, reference code CN{\textunderscore}00000013; M\'exico --
Consejo Nacional de Ciencia y Tecnolog\'\i{}a (CONACYT) No.~167733;
Universidad Nacional Aut\'onoma de M\'exico (UNAM); PAPIIT DGAPA-UNAM; The
Netherlands -- Ministry of Education, Culture and Science; Netherlands
Organisation for Scientific Research (NWO); Dutch national
e-infrastructure with the support of SURF Cooperative; Poland --
Ministry of Education and Science, grants
No.~DIR/WK/2018/11 and 2022/WK/12; National Science
Centre, grants No.~2016/22/M/ST9/00198,
2016/23/B/ST9/01635, 2020/39/B/ST9/01398, and 2022/45/B/ST9/02163;
Portugal -- Portuguese national funds and FEDER funds within Programa
Operacional Factores de Competitividade through Funda\c{c}\~ao para a Ci\^encia
e a Tecnologia (COMPETE); Romania -- Ministry of Research, Innovation
and Digitization, CNCS-UEFISCDI, contract no.~30N/2023 under Romanian
National Core Program LAPLAS VII, grant no.~PN 23\,21\,01\,02 and project
number PN-III-P1-1.1-TE-2021-0924/TE57/2022, within PNCDI III; Slovenia
-- Slovenian Research Agency, grants P1-0031, P1-0385, I0-0033, N1-0111;
Spain -- Ministerio de Econom\'\i{}a, Industria y Competitividad
(FPA2017-85114-P and PID2019-104676GB-C32), Xunta de Galicia (ED431C
2017/07), Junta de Andaluc\'\i{}a (SOMM17/6104/UGR, P18-FR-4314) Feder Funds,
RENATA Red Nacional Tem\'atica de Astropart\'\i{}culas (FPA2015-68783-REDT) and
Mar\'\i{}a de Maeztu Unit of Excellence (MDM-2016-0692); USA -- Department of
Energy, Contracts No.~DE-AC02-07CH11359, No.~DE-FR02-04ER41300,
No.~DE-FG02-99ER41107 and No.~DE-SC0011689; National Science Foundation,
Grant No.~0450696; The Grainger Foundation; Marie Curie-IRSES/EPLANET;
European Particle Physics Latin American Network; and UNESCO.
\end{sloppypar}


\begin{thebibliography}{99}
\footnotesize
\raggedright
\setlength{\itemsep}{0pt}
\def\vyp#1#2#3{\textbf{#1} (#2) #3} 


\bibitem{dipmodel}
  R.~Aloisio, V.~Berezinsky, P.~Blasi, A.~Gazizov, S.~Grigorieva, B.~Hnatyk, Astropart.\ Phys.\ \vyp{27}{2007}{76-91}

\bibitem{GZK1}
  K.~Greisen,  Phys.\ Rev.\ Lett.\ \vyp{16}{1966}{748}

\bibitem{GZK2}
 G.~T.~Zatsepin, V.~A.~Kuzmin, Zh.\ Eksp.\ Teor.\ Fiz.\ Pisma Red.\ \vyp{4}{1966}{144}

    
\bibitem{ICRC_convenga}
F.~Convenga, [Pierre Auger Coll.], PoS(ICRC2023)392 These proceedings

\bibitem{ICRC_anastasi}
G.~A.~Anastasi, [Pierre Auger Coll.], PoS(ICRC2023)343 These proceedings

\bibitem{augerNIM}
The Pierre Auger Collaboration, NIM \ A \vyp{798}{2015}{172}  

\bibitem{augerPrime}
A.~Aab et al., [Pierre Auger Coll.], [1604.03637] [astro-ph.IM].

\bibitem{ICRC_cataldi}
 G.~Cataldi, [Pierre Auger Coll.], PoS(ICRC2021)251  

\bibitem{ICRC_pawlosky}
J.~Pawlowsky, [Pierre Auger Coll.], PoS(ICRC2023)344 These proceedings

\bibitem{ICRC_dejesus}
J.~de~Jes\'us, [Pierre Auger Coll.], PoS(ICRC2023)267 These proceedings

\bibitem{ICRC_whisp}
 J.~C.~ Arteaga Vel\'azquez , [Pierre Auger coll.], PoS(ICRC2023)299

\bibitem{ICRC_nitz}
  D.~Nitz, [Pierre Auger coll.], PoS(ICRC2019)370

\bibitem{ICRC_sato}
 R.~ Sato,  [Pierre Auger coll.], PoS(ICRC2023)373 These proceedings
 
\bibitem{ICRC_schafer}
  C.M.Sch\"afer, [Pierre Auger coll.], PoS(ICRC2023)305 These proceedings

\bibitem{ICRC_bellido}
  J.~Bellido, [Pierre Auger coll.], PoS(ICRC2023)211 These proceedings

\bibitem{ICRC_harvey}
  V.~M.~Harvey, [Pierre Auger coll.], PoS(ICRC2023)300 These proceedings,  

\bibitem{ICRC_pallotta}
  J.~V.~Pallotta, [Pierre Auger coll.], PoS(ICRC2023)374 These proceedings  

\bibitem{ICRC_segreto}
  A.~Segreto, [Pierre Auger coll.], PoS(ICRC2023)276 These proceedings
  
\bibitem{ICRC_santos}  
  E.~Santos, [Pierre Auger coll.], PoS(ICRC2023)248 These proceedings
  
\bibitem{ICRC_langner}
  N.~Langner,  [Pierre Auger coll.], PoS(ICRC2023)371  These proceedings  

\bibitem{ICRC_hahn}
  S.~Hahn, [Pierre Auger coll.], PoS(ICRC2023)318 These proceedings  

\bibitem{ICRC_ellwanger}
  F.~Ellwanger, [Pierre Auger coll.], PoS(ICRC2023)275  These proceedings  

\bibitem{ICRC_zapparrata}
  O.~Zapparrata, [Pierre Auger coll.], PoS(ICRC2023)266  These proceedings  
  
\bibitem{ICRC_gcos}
  R.~Alves Batista et al. [GCOS], PoS(ICRC2023)281  These proceedings

\bibitem{auger_spec_prl}
  A.~Aab et al., [Pierre Auger Coll.], Phys.\ Rev.\ Lett.\ \vyp{125}{2020}{121106}
  
\bibitem{auger_spec_prd}
  A.~Aab et al., [Pierre Auger Coll.], Phys.\ Rev.\ D \vyp{102}{2020}{062005}
  
\bibitem{auger_spec_eurph}
  P.~Abreu et al., [Pierre Auger Coll.], Eur.\ Phys.\ J.\ C  \vyp{81}{2021}{966}

\bibitem{ICRC_novotny}
  V.~Novotn\`y,  [Pierre Auger Coll.], PoS(ICRC2021)324

\bibitem{ICRC_brichetto}
  G.~Brichetto Orquera, [Pierre Auger Coll.], PoS(ICRC2023)398 These proceedings

\bibitem{auger_mass1}
  J.~Abraham et al., [Pierre Auger Coll.], Phys.\ Rev.\ Lett.\ \vyp{104}{2010}{091101} [1002.0699]

\bibitem{auger_mass2}
  A.~Aab et al., [Pierre Auger Coll.], Phys.\ Rev.\ D\ \vyp{90}{2014}{122005} [1409.4809]

\bibitem{auger_mass3}  
  J.~Bellido, [Pierre Auger Coll.], PoS(ICRC2017)506

\bibitem{SD_comp}
  A.~Aab et al. [Pierre Auger Coll.], Phys.\ Rev.\ D\ \vyp{96}{2017}{122003}
  
\bibitem{ICRC_peixoto}
  C.~J.~T.~Peixoto,  [Pierre Auger Coll.], PoS(ICRC2019)440

\bibitem{DNN_erdmann}
 M.~Erdmann, J.~Glombitza, G.~Kasieczka, U.~Klemradt, World Scientific, 2021

\bibitem{DNN_general}
  S.~Hochreiter, J.~Schmidhuber, Neural Computation \vyp{9(8)}{1997}{1735–1780}

\bibitem{ICRC_mayotte}
  E.~W.~Mayotte, [Pierre Auger Coll.], PoS(ICRC2023)365 These proceedings
  
\bibitem{ICRC_fitoussi}
  T.~Fitoussi, [Pierre Auger Coll.], PoS(ICRC2023)319 These proceedings

\bibitem{ICRC_glombitza}
  J.~Glombitza, [Pierre Auger Coll.], PoS(ICRC2023)278 These proceedings

\bibitem{AERA_mass}
  B.~Pont, [Pierre Auger Coll.], EPJ\ Web\ Conf.\ \vyp{283}{2023}{02010}

\bibitem{HEAT_mass}
  A.~Yushkov, [Pierre Auger Coll.], PoS(ICRC2019)482

\bibitem{AUGER_combFit}
  A.~Abdul Halim et al., [Pierre Auger Coll.], JCAP \vyp{05}{2023}{024}

\bibitem{ICRC_olena}
  O.~Tkachenko, [Pierre Auger Coll.], PoS(ICRC2023)438 These proceedings

  \bibitem{ICRC_stadelmaier}
  M.~Stadelmaier, [Pierre Auger Coll.], PoS(ICRC2023)339 These proceedings

\bibitem{Auger_MuonFlucPRL} 
A.~Aab et al. [Pierre Auger Coll.], Phys.\ Rev.\ Lett.\ \vyp{126}{2021}{152002}

\bibitem{eposlhc}
  T.~Pierog, I.~Karpenko, J.~M.~Katzy, E.~Yatsenko and K.~Werner, Phys.\ Rev.\ C \ \vyp{92}{2015}{034906}

\bibitem{qgsjet}
  S.~Ostapchenko, Phys.\ Rev.\ D \ \vyp{83}{2011}{014018}
  
\bibitem{ICRC_gottowik}
  M.~Gottowik, [Pierre Auger Coll.], PoS(ICRC2023)345 These proceedings

\bibitem{ICRC_golup}
  G.~Golup, [Pierre Auger Coll.], PoS(ICRC2023)252 These proceedings
  
\bibitem{AUGER_anisAPJ}
   P.~Abreu et al., [Pierre Auger Coll.], Astrophys.\ J. \vyp{935}{2022}{170}  

\bibitem{ICRC_bister}
  T.~Bister, [Pierre Auger Coll.], PoS(ICRC2023)258 These proceedings
  
\bibitem{AUGER_largescale1}
   A.~Aab et al. [Pierre Auger Coll.], Science \vyp{357}{2017}{1266},

\bibitem{AUGER_largescale2}
  A.~Aab et al. [Pierre Auger Coll.], Astrophys.\ J.\ \vyp{868}{2018}{4}

\bibitem{ICRC_almeida}
  R.~M.~de Almeida, [Pierre Auger Coll.], PoS(ICRC2021)335

\bibitem{ICRC21_mayotte}
  E.~W.~Mayotte, [Pierre Auger Coll.], PoS(ICRC2021)321

\bibitem{ICRC_trimarelli}
  C.~Trimarelli, [Pierre Auger Coll.], PoS(ICRC2021)340  
  
\bibitem{auger_SHDM}
  P.~Abreu et al. [Pierre Auger Coll.], Phys.\ Rev.\ Lett.\ \vyp{130}{2023}{061001}

\bibitem{auger_liv}
  P.~Abreu et al. [Pierre Auger Coll.], [Pierre Auger Coll.], JCAP \vyp{01}{2022}{023}

\bibitem{ICRC_devito}
  E.~De Vito, [Pierre Auger Coll.], PoS(ICRC2023)1099 These proceedings

\bibitem{ICRC_yue}
  B.~Yue, [Pierre Auger Coll.], PoS(ICRC2023)1095 These proceedings
  
\bibitem{augerPhotons}
  P.~Abreu et al. [Pierre Auger Coll.], Universe \vyp{8}{2022}{579} 

\bibitem{ICRC_niechciol}
  M.~Niechciol, [Pierre Auger Coll.], PoS(ICRC2023)1488 These proceedings

\bibitem{ICRC_gonzalez}
  N.~Gonz\'alez, [Pierre Auger Coll.], PoS(ICRC2023)238 These proceedings

\bibitem{Auger_nu_JCAP}
  A. Aab et al. [Pierre Auger Coll.], JCAP \vyp{10}{2019}{022}

\bibitem{simprop}
R.~Aloisio, D.~Boncioli, A.~Di~Matteo, A.~F.~Grillo, S.~Petrera and F.~Salamida, JCAP \vyp{11}{2017}{009}
  
\bibitem{ICRC_petrucci}
  C. Petrucci, [Pierre Auger Coll.], PoS(ICRC2023)1520 These proceedings

\bibitem{ICRC_mussa}
  R.~Mussa, [Pierre Auger coll.], PoS(ICRC2023)372  These proceedings  

\bibitem{ICRC_colalillo}
  R.~Colalillo, [Pierre Auger coll.], PoS(ICRC2023)439  These proceedings  
  
\bibitem{ICRC_ghia}
  P.~L.~Ghia, [Pierre Auger coll.], PoS(ICRC2023)1616 These proceedings  

\bibitem{ICRC_sarmento}
  R.~Sarmento, [Pierre Auger coll.], PoS(ICRC2023)1611 These proceedings  

  
\end{thebibliography}
\end{document}